\documentclass[%
 reprint,
 superscriptaddress,
 amsmath,amssymb,
 aps,
 pre,
 floatfix,
]{revtex4-2}

\usepackage{amsmath}
\usepackage{amssymb}
\usepackage{mathtools}
\usepackage{graphicx}
\usepackage{dcolumn}
\usepackage{bm}
\usepackage{hyperref}
\usepackage[mathlines]{lineno}

\usepackage{xcolor}

\begin{document}

\title{Duplication--divergence-inspired model for growing scientific collaboration networks}

\author{Juan Garc\'ia-Castillo}
\affiliation{Instituto de F\'isica Interdisciplinar y Sistemas Complejos IFISC (CSIC-UIB), Palma de Mallorca, Spain}

\author{Beatriz Arregui-Garc\'ia}
\affiliation{Fondazione Bruno Kessler, Trento, Italy}

\author{Ant\`onia Tugores}
\affiliation{Instituto de F\'isica Interdisciplinar y Sistemas Complejos IFISC (CSIC-UIB), Palma de Mallorca, Spain}

\author{Jes\'us A. Moreno L\'opez\textsuperscript{\textdagger}}
\affiliation{Fondazione Bruno Kessler, Trento, Italy}

\author{Juan Fern\'andez-Gracia}
\affiliation{Instituto de F\'isica Interdisciplinar y Sistemas Complejos IFISC (CSIC-UIB), Palma de Mallorca, Spain}

\begin{abstract}
The structure of scientific collaboration networks emerges from how individual teams are put together: which members are kept, who leaves, and who joins. Within a single research institution these events can be followed paper by paper, yet they are rarely modelled explicitly. Here we introduce a growth model for the bipartite author--paper network inspired by the duplication--divergence mechanism: each new paper starts from the team of authors of an existing paper, chosen at random, which is then either replaced by an entirely new team of the same size, kept unchanged, or modified by removing some of its authors and adding new or previously active ones. Using rate equations, we show that the collaboration size relaxes as a power law in time towards a stationary shifted Poisson distribution, that the number of distinct authors grows linearly in time (asymptotically as $t\log t$ when authors are never removed), and that the uniform copying of papers induces an effective preferential attachment that produces a power-law distribution of papers per author. All predictions agree with numerical simulations. Finally, we calibrate the model on seventeen years of publications of a complex-systems research institute by simulation-based inference. We find that most papers arise from modifications of earlier collaborations, and that the calibrated model reproduces the observed growth of the number of authors and of the average productivity.
\end{abstract}

\maketitle

\section{Introduction}

Scientific research has become increasingly collaborative over the last decades \cite{repec:eee:respol:v:34:y:2005:i:3:p:259-285, wagner2005six, wuchty2007increasing}. The growing complexity of scientific problems, the need for multidisciplinary expertise, and the globalization of research have progressively increased both the size of research teams and the interconnectedness between scientists \cite{katz1997research, wu2019large}. As a consequence, collaboration networks have become one of the main tools for understanding how scientific knowledge is produced and how research communities evolve \cite{fortunato2018science}. Beyond describing patterns of co-authorship, these networks provide insight into productivity, interdisciplinarity, innovation, and the organization of scientific institutions.

Scientific collaboration networks have been extensively studied from the perspective of network science. Early work by Newman characterized the large-scale topological properties of co-authorship networks across multiple scientific disciplines, revealing common structural features such as heavy-tailed degree distributions, clustering, and small-world organization \cite{newman2001structure,newman2004coauthorship}. The heavy tail of the distribution of papers per author had in fact been documented long before as Lotka's law \cite{lotka1926frequency}, and cumulative-advantage processes were proposed early on to explain it \cite{simon1955class, price1976general}. Subsequent studies of the temporal evolution of these networks showed that preferential attachment explains the emergence of highly connected authors \cite{barabasi2002evolution}. Several generative models have since been proposed, including bipartite growth and preferential attachment models \cite{ramasco2004self, guillaume2006bipartite, peltomaki2006correlations} as well as team-based collaboration models \cite{guimera2005team, goldstein2005group, milojevic2014principles}. Empirical work has also shown that repeated collaborations are both frequent and consequential for scientific careers \cite{dahlander2013ties, petersen2015quantifying}. More recent approaches in the science of science have explored scientific impact \cite{wang2013quantifying, sinatra2016quantifying} and have used more refined representations such as higher-order interactions \cite{battiston2020networks, 10.3389/fphy.2023.1301994} and multilayer networks \cite{vasilyeva2021multilayer}. Together, these works have considerably improved our understanding of the macroscopic organization and evolution of scientific collaboration networks.

Nonetheless, the role of institutional organization in shaping collaboration has received far less attention. Existing models are primarily designed to reproduce the large-scale statistical properties of collaboration networks, rather than the mechanisms that generate collaborations within individual research institutions. As argued in Ref.~\cite{phillips2000inter}, collaboration and institutionalization reinforce each other: institutions provide the rules and resources that make collaboration possible, and collaborations in turn shape the institutions. Although scaling relations between the size of research institutions and their output have been characterized \cite{burghardt2021emergence}, few studies connect the elementary processes that generate collaborations inside an institution---the arrival of newcomers, the turnover of team members, repeated collaborations among established teams, or the formation of entirely new teams---with the structure of the resulting collaboration network.

In this work, we propose a mechanistic generative model for institutional collaboration networks inspired by the duplication--divergence paradigm. Duplication--divergence, in which a node is copied together with its links and the copy then loses some of them and gains new ones, was introduced to describe genome evolution by gene duplication \cite{ohno1970evolution}, and was later turned into growth models for protein interaction networks \cite{sole2002model, kim2002infinite, pastor2003evolving, vazquez2003modeling, ispolatov2005duplication, borrelli2025duplication}; closely related copying mechanisms were proposed for the growth of the World Wide Web \cite{kleinberg1999web, kumar2000stochastic, krapivsky2005network}. Here the duplicated object is the team of authors of a paper: a new paper starts from the authors of an existing one, who are then kept, modified or replaced. The model aims to explicitly represent elementary collaboration mechanisms: repeated collaborations, the turnover of team members, the arrival of new authors, and the creation of entirely new teams. Using rate equations, we derive the stationary distribution of authors per paper, the tail of the distribution of papers per author, and closed-form expressions for the time evolution of the number of distinct authors, the average number of papers per author, and the average number of authors per paper. Finally, we develop a simulation-based framework to infer the model parameters from empirical data, and we apply it to the publication record of a research institute, showing that these simple mechanisms reproduce to some extent the growth of the observed collaboration network, although not the full heterogeneity of its distributions.

\section{Duplication--divergence-inspired model} \label{sec:model}

\subsection{Definition of the model}

\begin{figure*}
    \centering
    \includegraphics[scale=0.45]{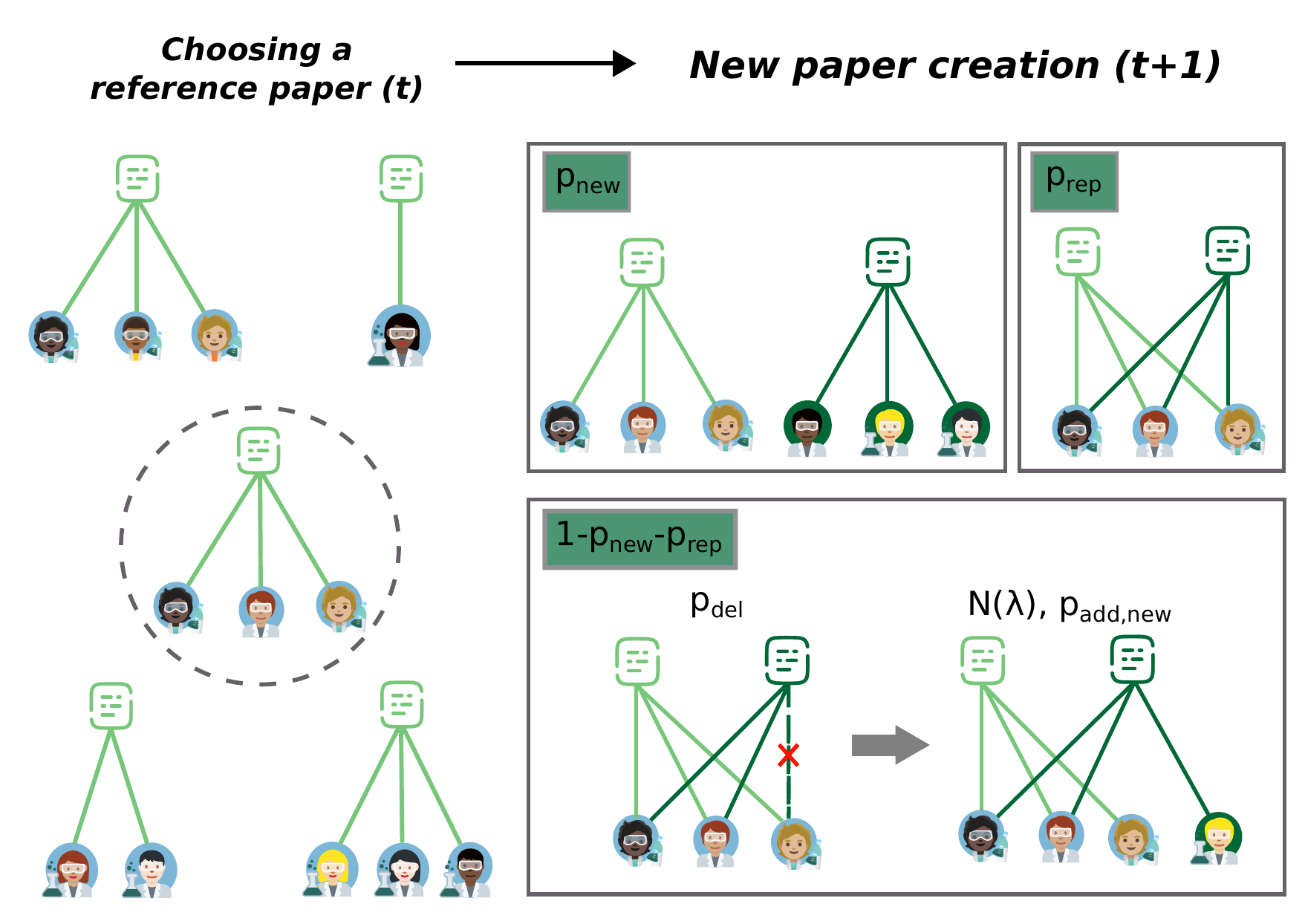}
    \caption{Scheme of the model. Existing papers are shown in light green and the newly generated paper is in dark green. Authors already present in the network before this step have a blue background, whereas authors introduced in this step have a green background.}
    \label{fig:model_scheme}
\end{figure*}

We model the growth of a bipartite (author--paper) collaboration network with a duplication--divergence mechanism, in which the duplicated object is the team of authors of a paper. Starting from an initial set of papers, at each time step one of the existing papers, called the reference paper, is drawn uniformly at random, and a new paper is generated by one of the following processes:
\begin{itemize}
    \item With probability $p_{\text{new}}$, a completely new team of authors, of the same size as the reference paper, writes a new paper.
    \item With probability $p_{\text{rep}}$, the authors of the reference paper write another paper together (a repeated collaboration).
    \item With probability $p_{\text{mod}}\coloneqq 1 - p_{\text{new}} - p_{\text{rep}}$, a modified version of the reference team writes the new paper:
    \begin{enumerate}
        \item One author of the reference paper, chosen uniformly at random, is always kept; each of the remaining authors is removed independently with probability $p_{\text{del}}$.
        \item A number $m$ of authors, drawn from a Poisson distribution of mean $\lambda$, is added to the team. Each added author is, with probability $p_{\text{add,new}}$, an author new to the network and, with probability $1-p_{\text{add,new}}$, an existing author chosen uniformly at random amongst the remaining authors in the network. If there is no such author, the addition is skipped.
    \end{enumerate}
\end{itemize}
These rules represent the arrival of entirely new teams ($p_{\text{new}}$), the persistence of established teams ($p_{\text{rep}}$), and the gradual turnover of teams ($p_{\text{mod}}$), which combines the departure of members from a collaboration ($p_{\text{del}}$) with the incorporation of new collaborators ($\lambda$), who may be newcomers to the network or already active in it ($p_{\text{add,new}}$). The process is iterated until the network reaches a prescribed number of papers. Figure~\ref{fig:model_scheme} sketches one step of the dynamics.

\subsection{Analytical characterization} \label{subsec:analytical}

We call the number of authors of a paper, $q$, its \textit{collaboration size}, and the number of papers of an author, $k$, the author's \textit{productivity}. Since one paper is created per time step, we identify time $t$ with the number of papers. Throughout the paper, we define $\alpha \coloneqq p_{\text{mod}}p_{\text{del}}$ and $q^* \coloneqq 1+\lambda/p_{\text{del}}$.

We derive rate equations for the ensemble averages of the average collaboration size $\langle q(t)\rangle$, the number of distinct authors $A(t)$ that appear in the network, and the average productivity $\langle k(t)\rangle$. We also give expressions for the stationary distribution of collaboration size $p_{\text{st}}(q)$ and the tail of the productivity distribution $p(k)$. Detailed derivations are given in Sec.~\ref{sec:expressions} of the Supplemental Material (SM), appended at the end of this document.

\subsubsection{Average collaboration size}

\begin{figure*}
    \centering
    \includegraphics[width=\textwidth]{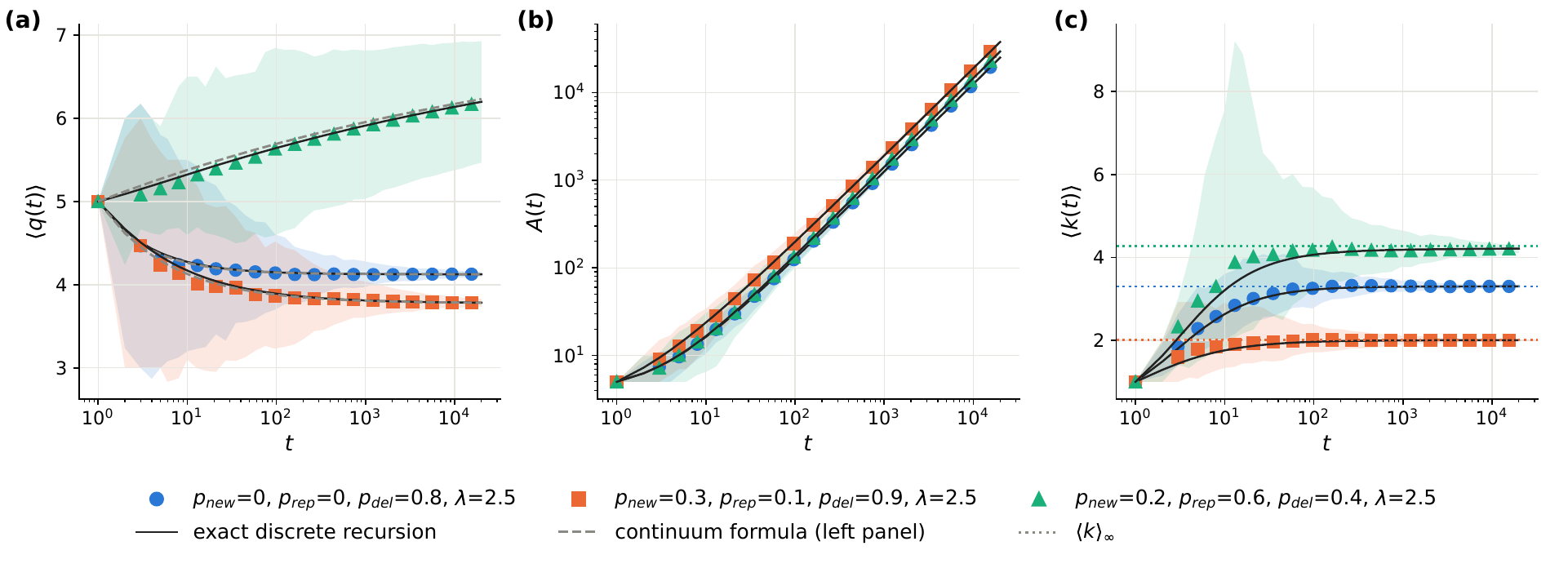}
    \caption{Time evolution of (a) the average collaboration size, $\langle q(t) \rangle$, (b) the number of distinct authors, $A(t)$, and (c) the average productivity, $\langle k(t) \rangle$, for the parameter sets of Table~\ref{tab:parameters-2}. Markers show the average over $100$ simulations and shaded areas the range of $95\%$ of the simulations. Solid lines show the exact solution of the discrete mean-field recursion (SM, Sec.~\ref{sec:appendix_average_authors_paper}), dashed lines the continuum formula of Eq.~\eqref{eq:solution_q_t}, and dotted lines the limit of Eq.~\eqref{eq:solution_k_t}.}
    \label{fig:time-evolution}
\end{figure*}

The new paper inherits the size of the reference in the new-team and repetition branches, whereas in the modification branch it keeps on average $1+(q-1)(1-p_{\text{del}})$ of the $q$ authors of the reference and gains $\lambda$ new ones. The total number of author--paper links, $S(t)=t\langle q(t)\rangle$, therefore grows on average by $\langle q\rangle - \alpha(\langle q\rangle -1) + p_{\text{mod}}\lambda$ per step. In the continuum limit, we obtain the following differential equation:
\begin{equation}
    \frac{d\langle q(t)\rangle}{dt} = \frac{p_{\text{mod}}}{t}\left[\lambda - p_{\text{del}}\left(\langle q(t)\rangle - 1\right)\right].
    \label{eq:rate_q}
\end{equation}
For $p_{\text{del}}>0$ its solution, starting from an initial network of $t_0$ papers, relaxes as a power law towards $q^*$,
\begin{equation} \label{eq:solution_q_t}
    \langle q(t)\rangle = q^* + \left[\langle q(t_0)\rangle - q^*\right]\left(\frac{t}{t_0}\right)^{-\alpha},
\end{equation}
whereas for $p_{\text{del}}=0$ the average size grows without bound,
\begin{equation}
    \langle q(t)\rangle = \langle q(t_0)\rangle + p_{\text{mod}}\lambda\log\left(\frac{t}{t_0}\right).
    \label{eq:solution-q_t-pdel-0}
\end{equation}
The underlying discrete recursion can also be solved exactly (SM, Sec.~\ref{sec:appendix_average_authors_paper}). The continuum expressions above are accurate for $t\gg t_0$ and differ only slightly from the exact solution at early times (Fig.~\ref{fig:time-evolution}).

\subsubsection{Number of distinct authors}

New authors enter the network in two ways: either as a whole new team, which happens with probability $p_{\text{new}}$ and introduces on average $\langle q(t)\rangle$ authors, or within a modified team, which happens with probability $p_{\text{mod}}$ and adds on average $\lambda$ authors, each of them new to the network with probability $p_{\text{add,new}}$. In the continuum limit, we again obtain a differential equation of the form:
\begin{equation}
    \frac{dA(t)}{dt} = p_{\text{new}}\langle q(t)\rangle + p_{\text{mod}}\,p_{\text{add,new}}\,\lambda .
    \label{eq:rate_A}
\end{equation}
For $p_{\text{del}}>0$, inserting Eq.~\eqref{eq:solution_q_t} gives a linear growth with a power-law correction at early times,
\begin{equation}\label{eq:solution_A_t}
\begin{aligned}
    A(t) = {}& A(t_0) + a\,(t-t_0) \\
   &+ p_{\text{new}}\left[\langle q(t_0)\rangle-q^*\right]\frac{t_0^{\alpha}}{1-\alpha}\left(t^{1-\alpha}-t_0^{1-\alpha}\right),
\end{aligned}
\end{equation}
with asymptotic growth rate $a \coloneqq p_{\text{new}}q^* + p_{\text{mod}}p_{\text{add,new}}\lambda$.
For $p_{\text{del}}=0$, instead,
\begin{equation}
\begin{aligned}
A(t) ={}& A(t_0) + \left[p_{\text{new}}\langle q(t_0)\rangle + p_{\text{mod}}p_{\text{add,new}}\lambda\right](t-t_0) \\
& + p_{\text{new}}p_{\text{mod}}\lambda \left[ t\log\left(\frac{t}{t_0}\right) -(t-t_0) \right],
\end{aligned}
\label{eq:solution_A_t-pdel-0}
\end{equation}
so that the number of authors grows asymptotically as $t\log t$ whenever $p_{\text{new}}>0$, and linearly only if $p_{\text{new}}=0$.

\subsubsection{Average productivity}

Since the total number of author--paper links can be counted from either side of the bipartite network, $t\langle q(t)\rangle = A(t)\langle k(t)\rangle$. Having expressions for both $\langle q(t)\rangle$ and $A(t)$, we obtain the average productivity as
\begin{equation}
    \langle k(t)\rangle = \frac{t\,\langle q(t)\rangle}{A(t)} .
    \label{eq:av_productivity}
\end{equation}
Equation~\eqref{eq:av_productivity} gives the full time evolution of $\langle k(t)\rangle$; here we report only its long-time limits.

For $p_{\text{del}}>0$ the average productivity converges to
\begin{equation}
    \lim_{t\to\infty}\langle k(t)\rangle = \frac{q^*}{a} = \frac{p_{\text{del}} + \lambda}{p_{\text{new}}(p_{\text{del}} + \lambda) + p_{\text{mod}}\,p_{\text{add,new}}\,\lambda\, p_{\text{del}}} .
    \label{eq:solution_k_t}
\end{equation}
For $p_{\text{del}}=0$ and $p_{\text{new}}>0$, both $\langle q\rangle$ and $A/t$ grow logarithmically, and
\begin{equation}
    \lim_{t\to\infty}\langle k(t)\rangle = \frac{1}{p_{\text{new}}},
    \label{eq:solution_k_t-pdel-0}
\end{equation}
a limit that is approached slowly, with corrections of order $1/\log t$; if moreover $p_{\text{new}}=0$, $\langle k(t)\rangle$ grows as $\log t$.

\subsubsection{Stationary distribution of collaboration size}

For $p_{\text{del}}>0$, the distribution of collaboration size converges at large times to a 1-shifted Poisson distribution with parameter $\Lambda = \lambda/p_{\text{del}}$. We obtain this result by calculating the stationary distribution with the probability generating function method (SM, Sec.~\ref{sec:appendix_stationary_authors_papers}):
\begin{equation}
    p_{\text{st}}(q)= e^{-\Lambda}\,\frac{\Lambda^{q-1}}{(q-1)!}, \qquad q\geq 1,
    \label{eq:p_q-stationary}
\end{equation}
whose mean $q^*$ coincides with the limit of Eq.~\eqref{eq:solution_q_t}.

This result has a simple interpretation: independent deletions thin the Poisson-distributed number of non-protected authors, and the Poisson additions restore it, so a Poisson distribution of mean $\Lambda$ is preserved when $\Lambda p_{\text{del}}=\lambda$. Note that the relaxation towards $p_{\text{st}}$ is slow, with the same exponent $\alpha$ as in Eq.~\eqref{eq:solution_q_t}. For $p_{\text{del}}=0$, there is no stationary distribution, since the average collaboration size grows without bound.

\subsubsection{Tail of the productivity distribution}

Because the reference paper is drawn uniformly at random, an author with $k$ papers belongs to it with probability $k/t$: the uniform copying of papers acts as an effective linear preferential attachment on authors, i.e., a Simon--Yule cumulative-advantage process \cite{simon1955class, price1976general}. An author of the reference paper is always kept in a repetition, and is kept in a modification with probability $1-p_{\text{del}}(1-1/q)$, where $1/q$ is the probability of being the protected author. Averaging over the papers of an author, whose sizes are size-biased, gives $\sum_{j\ni i}1/q_j \simeq k_i/\langle q\rangle$, so that for long times
\begin{equation}
    \frac{dk_i}{dt} \simeq B\,\frac{k_i}{t} + \frac{c}{t},
    \label{eq:rate_k}
\end{equation}
with
\begin{equation}
    B = p_{\text{rep}} + p_{\text{mod}}\left[1 - p_{\text{del}}\left(1-\frac{1}{q^*}\right)\right],
    \label{eq:B}
\end{equation}
where the constant $c = p_{\text{mod}}\lambda(1-p_{\text{add,new}})/a$ accounts for existing authors that are added to a team. Since authors enter the network at a constant rate, an author born at time $s$ has $k_i(t)\sim(t/s)^{B}$, and the productivity distribution develops a power-law tail $p(k)\sim k^{-\gamma}$ with
\begin{equation}
\gamma = 1+\frac{1}{B} = 1 + \frac{1}{p_{\text{rep}} + p_{\text{mod}}\left(1 - \dfrac{\lambda\,p_{\text{del}}}{\lambda+p_{\text{del}}}\right)} .
\label{eq:gamma}
\end{equation}

\begin{figure*}
    \centering
    \includegraphics[width=\textwidth]{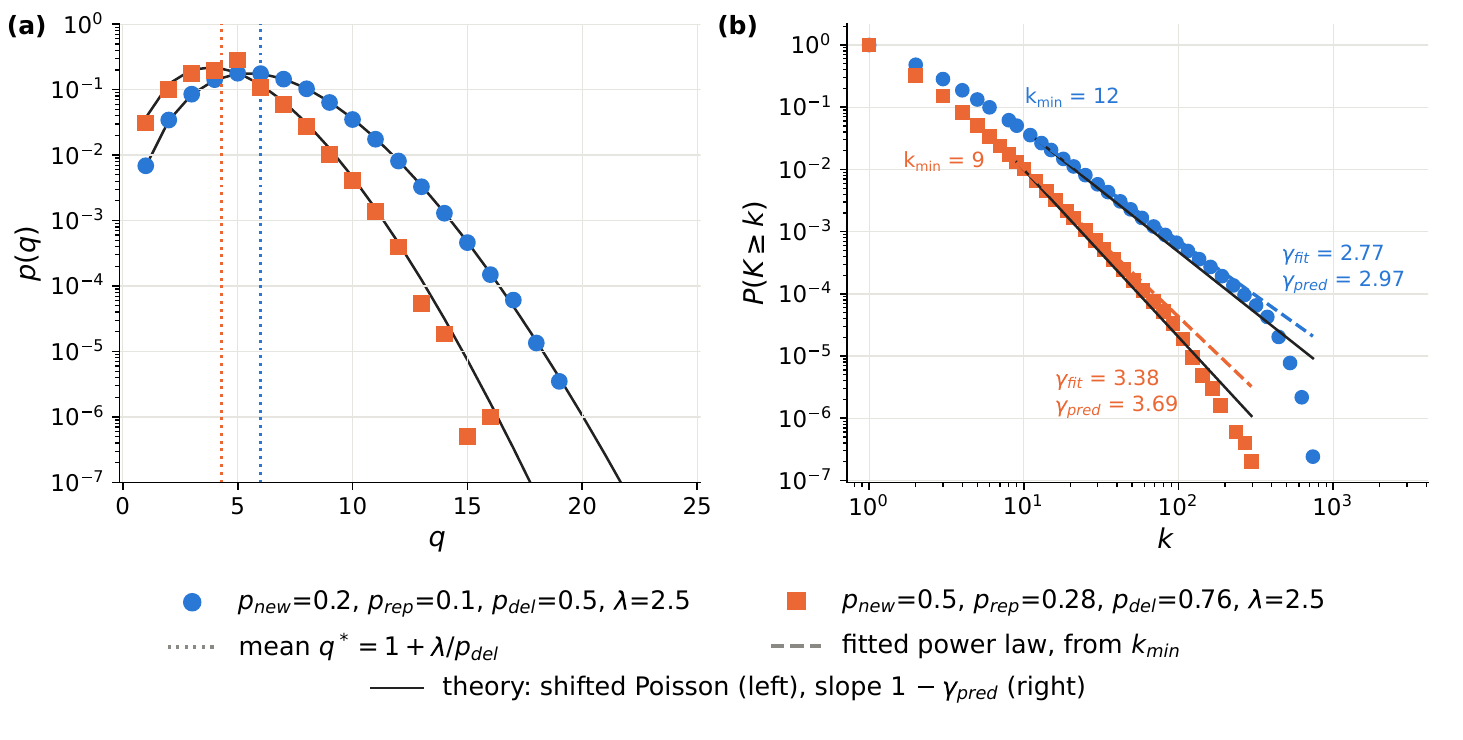}
    \caption{Distributions of (a) collaboration size, $p(q)$, and (b) productivity (complementary cumulative distribution) $P(K\geq k)$. In both panels, markers show the distribution pooled over $100$ simulations of $20\,000$ papers for the parameter sets of Table~\ref{tab:parameters}. Solid lines show both the shifted Poisson distribution of Eq.~\eqref{eq:p_q-stationary} in (a) as well as the slope $1-\gamma$ predicted by Eq.~\eqref{eq:gamma} in (b). Dashed lines in (b) are the fitted power-laws drawn by starting from the fitted cut-off $k_{\min}$.}
    \label{fig:theoretical-distributions}
\end{figure*}

Since $B\leq p_{\text{rep}}+p_{\text{mod}} = 1-p_{\text{new}}$, the exponent satisfies $\gamma\geq 2$. Figure~\ref{fig:theoretical-distributions} compares Eqs.~\eqref{eq:p_q-stationary} and~\eqref{eq:gamma} with simulations (Sec.~\ref{subsec:validation}).

\subsection{Numerical validation}
\label{subsec:validation}

We validate the results of Sec.~\ref{subsec:analytical} against $100$ independent simulations of the model, each grown up to $20\,000$ papers, which allows us to estimate the fluctuations across realizations and to reach the long-time regime when possible. All simulations start from a single paper with $5$ authors, i.e., $t_0 = 1$, $\langle k(t_0)\rangle = 1$  and $A(t_0)=\langle q(t_0)\rangle=5$. Here we focus on the case $p_{\text{del}}>0$; the case $p_{\text{del}}=0$ is validated in SM, Sec.~\ref{sec:appendix_verification_del_0}.

\subsubsection{Time evolution of $\langle q(t) \rangle$, $A(t)$, and $\langle k(t)\rangle$}

To validate the time-dependent results, we use the parameter sets listed in Table~\ref{tab:parameters-2}, which cover high, intermediate, and low values of $p_{\text{mod}}$. For the high and intermediate sets, the parameters are chosen to ensure that $\langle q(t) \rangle$ converges (see SM, Sec.~\ref{sec:appendix_convergence_authors_paper}). For the low $p_{\text{mod}}$ set, $\langle q(t)\rangle$ does not converge within the simulated time unless the system starts in the near vicinity of the stationary regime; we include it to test the formulas also far from stationarity.

\begin{table}[t]
\caption{Parameter sets used to validate the time evolution of $A(t)$, $\langle k(t)\rangle$ and $\langle q(t) \rangle$.}
\label{tab:parameters-2}
\begin{ruledtabular}
\begin{tabular}{ccccc}
$p_{\text{new}}$ & $p_{\text{rep}}$ & $p_{\text{del}}$ & $\lambda$ & $p_{\text{add,new}}$ \\
\colrule
0   & 0   & 0.8   & 2.5 & 0.5 \\
0.3 & 0.1 & 0.9   & 2.5 & 0.5 \\
0.2 & 0.6 & 0.4 & 2.5 & 0.5 \\
\end{tabular}
\end{ruledtabular}
\end{table}

Fig.~\ref{fig:time-evolution}(a) shows that the average collaboration size relaxes towards $q^*=1+\lambda/p_{\text{del}}$ as predicted by Eq.~\eqref{eq:solution_q_t} for the high and intermediate $p_{\text{mod}}$ sets. Theory and simulations also agree for the low $p_{\text{mod}}$ set, which has not yet converged to the stationary state.

Figure~\ref{fig:time-evolution}(b) shows that the number of distinct authors agrees with Eq.~\eqref{eq:solution_A_t}. After a short transient, $A(t)$ grows linearly with time, which is the property used to derive Eq.~\eqref{eq:gamma}.

Lastly, Fig.~\ref{fig:time-evolution}(c) shows that Eq.~\eqref{eq:av_productivity} captures in detail the time evolution of the average productivity, including its long-time limit, Eq.~\eqref{eq:solution_k_t}, which equals $3.30$, $2.01$ and $4.26$ for the three parameter sets.

\subsubsection{Distributions of collaboration size and productivity}

To validate Eqs.~\eqref{eq:p_q-stationary} and~\eqref{eq:gamma}, we use the parameter sets of Table~\ref{tab:parameters}, chosen to ensure the predicted exponent $\gamma$ falls in the range of values reported for empirical collaboration networks \cite{lotka1926frequency, newman2001structure, barabasi2002evolution}.

\begin{table}[h]
\caption{Parameter sets used to validate the stationary distribution of collaboration size, Eq.~\eqref{eq:p_q-stationary}, and the exponent of the productivity distribution, Eq.~\eqref{eq:gamma}.}
\label{tab:parameters}
\begin{ruledtabular}
\begin{tabular}{ccccc}
$p_{\text{new}}$ & $p_{\text{rep}}$ & $p_{\text{del}}$ & $\lambda$ & $p_{\text{add,new}}$ \\
\colrule
0.2  & 0.1  & 0.5  & 2.5 & 0.5 \\
0.5  & 0.28 & 0.76 & 2.5 & 0.5 \\
\end{tabular}
\end{ruledtabular}
\end{table}

Figure~\ref{fig:theoretical-distributions}(a) shows that, for both parameter sets, the simulated distribution of collaboration size agrees with the shifted Poisson distribution of Eq.~\eqref{eq:p_q-stationary}.  The simulated productivity distributions [Fig.~\ref{fig:theoretical-distributions}(b)] follow the slope predicted by Eq.~\eqref{eq:gamma} over a wide range of $k$, up to a cut-off at large $k$ due to the finite simulation time. The exponents are fitted using the discrete maximum-likelihood method of Ref.~\cite{clauset2009power}, with the start of the fit range, $k_{\min}$, chosen by minimizing the Kolmogorov--Smirnov distance between the data above it and the fitted power law. The fitted exponents, $\gamma = 2.77\pm0.01$ and $\gamma=3.38\pm0.02$, are close to, but smaller than, the predictions of Eq.~\eqref{eq:gamma}, $\gamma=2.97$ and $\gamma=3.69$. We attribute the difference to finite-size effects. Equation~\eqref{eq:gamma} gives the exponent of the tail for $k\to\infty$, while at moderate $k$ the mean-field solution itself is not yet a pure power law, with the distribution taking the form $P(K\geq k)\propto(k+c/B)^{-1/B}$ (SM, Sec.~\ref{sec:appendix_tail_papers_authors}). This means a flatter slope than $k^{-1/B}$ when $k$ is not much larger than $c/B$. In addition, Eq.~\eqref{eq:gamma} neglects the fluctuations of the sizes of the papers of an author around the mean-field approximation $\sum_{j\ni i}1/q_j \simeq k_i/\langle q\rangle$, and the early transients of $\langle q(t)\rangle$ and $A(t)$.

\section{Data} \label{sec:data}

We use the publication record of the Institute for Cross-Disciplinary Physics and Complex Systems (IFISC), a research institute devoted to the study of complex systems. The records are taken from the internal publication database of the institute and include all the journal articles published between 2007 and 2023, both years included, together with their external co-authors. We only keep papers with at most $25$ authors, since very large collaborations follow dynamics different from the ones that our model aims to capture. The resulting dataset contains $1317$ papers written by $1872$ distinct authors. Author names were manually disambiguated by checking name variants and merging entries referring to the same individual, including those using full given names or initials.

From these records we build a bipartite author--paper network by linking each article to its authors. Papers are ordered by publication date, so that the number of papers plays the role of time, as in the model. In the specific case in which the publication dates coincide, papers are then ordered alphabetically by title.

\section{Parameter inference} \label{sec:estimators}

\subsection{Summary statistics}

To compare the model with data, we define five statistics, one for each model parameter, that can be computed in the same way on the observed corpus and on any simulated realization. Each statistic is designed to measure, as far as the data allow, the corresponding parameter or a closely related quantity. For this, we impose two basic assumptions: any previously published paper is an equally plausible reference for a new one, irrespective of its age, and every paper can act as a reference as soon as it is published.

We take then the first paper as the seed and process every subsequent paper in chronological order and compare it with all papers published before it. We denote by $C$ the author set of the paper under study, by $\mathcal{C}_0$ the family of distinct author sets among all previously processed papers, and by $\mathcal{A}_0 \coloneqq \bigcup_{P \in \mathcal{C}_0} P$ the set of all authors that have appeared so far. Each paper is then classified as follows:
\begin{enumerate}
    \item \textit{new} if it shares no author with any previous paper, $C \cap \mathcal{A}_0 = \emptyset$;
    \item \textit{repeated} if its author set coincides with that of a previous paper, $C\in\mathcal{C}_0$;
    \item \textit{modified} otherwise, that is, if it shares at least one of its authors with a previous paper.
\end{enumerate}

Denoting by $N_{\text{new}}$, $N_{\text{rep}}$ and $N_{\text{mod}}$ the number of papers in each class, and by $N = N_{\text{new}} + N_{\text{rep}} + N_{\text{mod}}$ the number of processed papers (all papers except the seed), we define
\begin{equation}
    s_{\text{new}} = \frac{N_{\text{new}}}{N}, \qquad s_{\text{rep}} = \frac{N_{\text{rep}}}{N}.
\end{equation}

All remaining statistics are computed over the $N_{\text{mod}}$ modified papers, and sums indexed by ``$\text{mod } C$'' run over the author set $C$ of the aforementioned $N_{\text{mod}}$ papers. Since we established earlier that every paper is equally likely to be the reference of a modified one, we consider all previous papers sharing at least one author with it, $\mathcal{O}(C) = \left\{ P_j \in \mathcal{C}_0 : P_j \cap C \neq \emptyset \right\}$, and give each pair $(C, P_j)$ the weight $1/n_C$, with $n_C = |\mathcal{O}(C)|$, so that every modified paper contributes equally. For each pair we count both the authors of $P_j$ absent from $C$, $|P_j \setminus C|$, and the authors of $C$ absent from $P_j$, $|C\setminus P_j|$.

The deletion statistic is hence the weighted fraction of removed authors among those exposed to deletion,
\begin{equation}
    s_{\text{del}} = \frac{\displaystyle\sum_{\text{mod } C}\ \frac{1}{n_C}\sum_{P_j \in \mathcal{O}(C)} |P_j \setminus C|}{\displaystyle\sum_{\text{mod } C}\ \frac{1}{n_C}\sum_{P_j \in \mathcal{O}(C)} (|P_j|-1)} ,
\end{equation}
where the denominator uses $|P_j|-1$ because one author of the reference is always kept. Single-author references carry no information and are excluded from both sums.

Likewise, the addition statistic is the weighted average number of added authors per modified paper,
\begin{equation}
    s_{\text{add}} = \frac{1}{N_{\text{mod}}}\sum_{\text{mod } C}\ \frac{1}{n_C} \sum_{P_j \in \mathcal{O}(C)} |C \setminus P_j| ,
\end{equation}
and the newcomer statistic is the weighted fraction of added authors who had never appeared before,
\begin{equation}
  s_{\text{add,new}} = \frac{\displaystyle\sum_{\text{mod } C}\ \frac{1}{n_C} \sum_{P_j \in \mathcal{O}(C)}\ \sum_{u \in C \setminus P_j} \nu(u)}{\displaystyle\sum_{\text{mod } C}\ \frac{1}{n_C} \sum_{P_j \in \mathcal{O}(C)} |C \setminus P_j|} ,
\end{equation}
where $\nu(u)=1$ if author $u$ does not belong to $\mathcal{A}_0$ and $\nu(u)=0$ otherwise.

The statistics $s_{\text{new}}$ and $s_{\text{rep}}$ directly measure the corresponding probabilities $p_{\text{new}}$ and $p_{\text{rep}}$. The other three, however, are not unbiased estimators of the model parameters. For instance, comparing a modified paper with all overlapping previous papers, rather than with its single, actual reference, inflates $|P_j\setminus C|$, and hence $s_{\text{del}}$, with respect to $p_{\text{del}}$ (see SM, Sec. \ref{sec:appendix_bias}).

\subsection{Calibration}
\label{sec:calibration}

We begin by defining our parameter vector $\Theta=(p_{\text{new}}, p_{\text{rep}}, p_{\text{del}}, \lambda, p_{\text{add,new}})$ and six summary statistics $s_i$: the five statistics defined above and the number of distinct authors $s_A$. This additional statistic ensures that the simulated networks have a realistic number of authors. For every simulated dataset, we take the papers of the first year (2007) as a seed, and we grow the network up to the size of the observed corpus, $1317$ papers. The simulated statistics are thus computed on data of the same size, from the same initial condition and with the same classification procedure as the observed data.

We then look for the parameter vector that minimizes the following distance between the simulated and observed statistics:
\begin{equation}
d(\Theta) = \sum_{i=1}^{6} \frac{\left| \bar s_i(\Theta) - s_i^{\text{obs}} \right|}{\sigma_i},
\label{eq:fitness}
\end{equation}
where $\bar s_i(\Theta)$ is the average of $s_i$ over independent simulations of the model and $\sigma_i$ is the standard deviation of $s_i$ across simulations, that is, the intrinsic noise of the simulator on that statistic. Each term of Eq.~\eqref{eq:fitness} is thus measured in units of noise. The parameters are restricted to $p_{\text{new}}, p_{\text{rep}}, p_{\text{del}}, p_{\text{add,new}} \in [0,1]$ and $\lambda \in [0,5]$, and we impose $p_{\text{new}}+p_{\text{rep}}\leq 1$, in accordance with the rules of the model.

We first locate the best fit with a global search: the distance is evaluated on $1000$ points of a Latin hypercube design over the parameter box~\cite{mckay1979comparison}, with $20$ simulations per point. The best points are afterwards refined with the Nelder--Mead algorithm~\cite{nelder1965simplex}. We then approximate the posterior distribution of $\Theta$ with an approximate Bayesian computation scheme with sequential Monte Carlo (ABC-SMC)~\cite{beaumont2002approximate, toni2008approximate}, starting from a uniform prior over the parameter box. A population of $1000$ particles is propagated through $20$ generations of decreasing tolerance $\varepsilon$, for a total of about $4.2\times10^{5}$ simulations. All numerical settings are given in SM, Sec.~\ref{sec:appendix_calibration}.

Table~\ref{tab:results} reports the summary statistics computed on the data and on simulations at the best fit, the best fit, and the posterior medians with $95\%$ credible intervals. According to these values, about $3\%$ of the papers of the institute are written by an entirely new team, about $19\%$ repeat an earlier team exactly, and the remaining $78\%$ modify an earlier team. In a modification, each non-protected author leaves with probability $p_{\text{del}}\approx0.55$, and about two authors join, each of them new to the network with a probability of about $80\%$. At the best fit, five of the six statistics of the data lie within the $95\%$ range of the simulations, and the deletion statistic lies just below it (Table~\ref{tab:results} and SM, Fig.~\ref{fig:best_fit_summaries}).

\begin{table*}[t]
\caption{Summary statistics and inferred parameters. Left block: each summary statistic computed on the observed corpus, and its $95\%$ range over $30$ simulated corpora at the best fit. Right block: the model parameter associated with it, at the best fit of the global search and as ABC-SMC posterior median with its $95\%$ credible interval (CI). A statistic is not an estimate of its parameter (see text and SM, Sec.~\ref{sec:appendix_bias}).}
\label{tab:results}
\begin{ruledtabular}
\begin{tabular}{lcc|lccc}
\multicolumn{3}{c|}{Summary statistic} & \multicolumn{4}{c}{Model parameter} \\
 & Data & Simulated at best fit ($95\%$) &  & Best fit & Median & $95\%$ CI \\
\colrule
$s_{\text{new}}$     & 0.038 & [0.029, 0.044] & $p_{\text{new}}$     & 0.031 & 0.031 & [0.013, 0.048] \\
$s_{\text{rep}}$     & 0.193 & [0.185, 0.215] & $p_{\text{rep}}$     & 0.187 & 0.181 & [0.143, 0.223] \\
$s_{\text{del}}$     & 0.910 & [0.911, 0.925] & $p_{\text{del}}$     & 0.555 & 0.539 & [0.448, 0.652] \\
$s_{\text{add}}$     & 3.263 & [3.11, 3.36]   & $\lambda$            & 2.01  & 1.98  & [1.72, 2.32] \\
$s_{\text{add,new}}$ & 0.515 & [0.492, 0.520] & $p_{\text{add,new}}$ & 0.793 & 0.814 & [0.709, 0.913] \\
$s_A$                & 1872  & [1809, 1934]   & ---                  & ---   & ---   & --- \\
\end{tabular}
\end{ruledtabular}
\end{table*}

The calibrated parameters imply a stationary average collaboration size $q^*=1+\lambda/p_{\text{del}}\approx4.6$ [$95\%$ credible interval $4.38$--$5.02$], a productivity exponent $\gamma\approx2.6$ [$2.46$--$2.76$] and a long-time average productivity $\langle k\rangle_\infty\approx3.3$ [$3.09$--$3.54$], from Eqs.~\eqref{eq:gamma} and~\eqref{eq:solution_k_t}. Each credible interval is obtained by evaluating the corresponding expression on every particle of the ABC-SMC posterior, which accounts for the correlations between parameters.

\subsection{Identifiability of the inference}
\label{sec:identifiability}

Some combinations of the parameters of a model are usually much better constrained by the data than others~\cite{Gutenkunst2007, transtrum2015perspective}. We assess how well our data determine the parameters with three complementary analyses. First, profiles of the objective function, in the spirit of the profile likelihood~\cite{raue2009structural}, search the whole parameter box for competing minima. Second, the Fisher information at the best fit quantifies locally how tightly each combination of parameters is constrained~\cite{rao1945information, cramer1946mathematical}. Third, inference on synthetic corpora tests whether the whole procedure recovers known parameters~\cite{cook2006validation, talts2018validating}. Details and supporting figures are given in SM, Secs.~\ref{sec:appendix_multimodality}--\ref{sec:appendix_recovery}.

\paragraph{Profiles.} We fix one parameter at a time and re-optimize the others with the full distance of Eq.~\eqref{eq:fitness}. Every profile has a single, well-defined minimum at the best fit, and moving away from it raises the distance by one to two orders of magnitude (SM, Fig.~\ref{fig:profiles}). The objective function thus has no competing minima and no flat directions over the parameter box.

\paragraph{Sloppiness.} Sloppiness analysis identifies the directions in parameter space along which the cost function remains nearly unchanged: along them, different parameter sets fit the data almost equally well, so the data cannot determine the corresponding combinations of parameters~\cite{Gutenkunst2007, transtrum2015perspective}. In a sloppy model these soft directions are orders of magnitude less constrained than the stiff ones, so that individual parameters are poorly determined, although predictions that depend mainly on the stiff combinations can remain reliable. We quantify this with the Fisher information $F$ at the best fit, taken with respect to the logarithms of the parameters (SM, Sec.~\ref{sec:appendix_sloppiness}). Near the best fit, the parameter sets that the data cannot distinguish from it within one standard deviation form an ellipsoid. Its axes are the eigenvectors of $F$, each a weighted sum of the logarithms of the parameters, and the half-length of each axis is $1/\sqrt{\mu}$, with $\mu$ the corresponding eigenvalue: moving the parameters along an axis by a relative amount $1/\sqrt{\mu}$ shifts the average summary statistics of the simulations by one standard deviation of the simulator's noise. This half-length is also the smallest relative standard deviation with which any estimator can determine the corresponding combination of parameters (Cram\'er--Rao bound~\cite{rao1945information, cramer1946mathematical}).

The longest axis of the ellipsoid is only about ten times the shortest, so the eigenvalues span about two orders of magnitude (condition number $\kappa=\mu_{\max}/\mu_{\min}\approx10^{2}$), whereas in typical sloppy models they span many more~\cite{Gutenkunst2007, transtrum2015perspective}. No combination of parameters is thus left undetermined by the data, and the model is only mildly sloppy. The two stiffest axes correspond to the features of the collaboration process that the corpus records most directly. The first, dominated by $p_{\text{add,new}}$, is constrained to within $1.5\%$, because newcomers directly drive the number of distinct authors, one of the most precisely measured statistics. The second is the ratio $\lambda/p_{\text{del}}$, which sets the stationary collaboration size $q^*=1+\lambda/p_{\text{del}}$ at which additions balance deletions, and is determined to about $\pm2\%$. This precision holds within the model: $q^*$ is fixed indirectly, through the deletion and addition statistics, and the model overestimates the observed average collaboration size (Sec.~\ref{sec:analysis}). The loosest axis is $p_{\text{new}}$ alone at $\pm15\%$ because papers by entirely new teams are rare---about $50$ in the corpus---so their share carries a large relative counting noise. In absolute terms, however, this is only $\pm0.005$, and $p_{\text{new}}$ barely enters the main predictions of the model.

These results explain the intervals of Table~\ref{tab:results}. The interval of $p_{\text{new}}$ is wide relative to its value because $p_{\text{new}}$ lies along the loosest direction of the ellipsoid, although in absolute terms it remains small. The intervals of $p_{\text{del}}$ and $\lambda$ are wide for a different reason: the data mainly constrain their ratio, and with it the stationary collaboration size $q^*$, but not each parameter separately. As long as $p_{\text{del}}$ and $\lambda$ increase or decrease together, $q^*$ does not change and neither does the fit, so both parameters can slide along this direction, and $p_{\text{add,new}}$ slides with them in the opposite direction (SM, Tables~\ref{tab:fisher} and~\ref{tab:correlations}). The individual values of these three parameters should therefore be read with caution, whereas the quantities on which our results rest, $q^*$, $\langle k\rangle_\infty$ and $\gamma$, remain tightly constrained (see also Sec.~\ref{sec:discussion}).

\paragraph{Recovery.} We test the whole procedure on synthetic data. For four parameter sets with known values---the best fit and three deliberately different sets---we simulate $100$ corpora each, of the same size as the observed one, and infer their parameters as if they were the data (SM, Fig.~\ref{fig:recovery}). The inference uses the same six summary statistics as for the data, and all five parameters are estimated. This tests whether the calibration method recovers the parameters in different regions of the parameter space.

At the best fit, the root-mean-square errors of the posterior medians are $0.004$, $0.029$, $0.024$, $0.021$ and $0.08$ for $p_{\text{new}}$, $p_{\text{rep}}$, $p_{\text{del}}$, $p_{\text{add,new}}$ and $\lambda$, within $0.8$--$2.6$ times the Cram\'er--Rao bounds, and the $95\%$ credible intervals contain the true value in $93\%$--$100\%$ of the corpora in $18$ of the $20$ combinations of parameter and set. The weak spots are a slight underestimation of $p_{\text{rep}}$, by about $0.03$, and, for the set with the highest turnover, a slight overestimation of $p_{\text{del}}$ and $\lambda$ along the ridge described above. The full ABC-SMC pipeline, run on three synthetic corpora generated at the best fit, contained all $15$ true parameter values within its $95\%$ intervals.

Together, these analyses show that all five parameters are identified by the six statistics, and that, if the data were generated by the model, our procedure would recover the parameters with well-calibrated uncertainties at the size of our corpus.

\section{Model versus data} \label{sec:analysis}

To test the calibrated model beyond the statistics used in its calibration, we compare the observed and simulated distributions of collaboration size and productivity. We measure their difference with the Earth Mover's Distance (EMD)~\cite{rubner2000earth} and judge it against the EMDs between replicas of the model and the model mean, that is, with a parametric bootstrap.

We also compare the time evolution of the three mean-field quantities---average collaboration size, number of distinct authors and average productivity---in the data and in the model. To do this, we simulate corpora with parameters drawn from the posterior and at the best fit, and compare them with the value of each quantity in the data at the end of every year. We use the cumulative number of publications as time, so that the time scales of data and model match.

\subsection{Time evolution of $\langle q(t) \rangle$, $A(t)$ and $\langle k(t) \rangle$ }

\begin{figure*}[t]
    \centering
    \includegraphics[width=\textwidth]{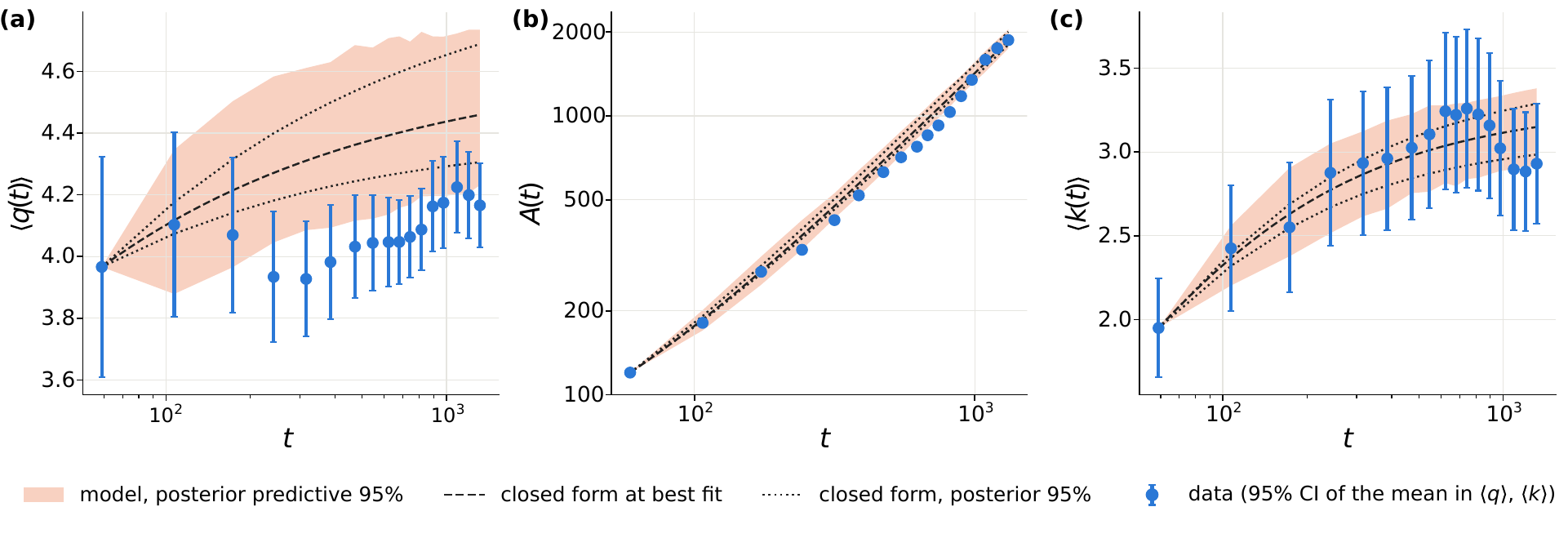}
    \caption{(a) Cumulative average collaboration size, (b) number of distinct authors and (c) cumulative productivity as a function of the number of papers, with one point per year. Blue markers: data; in (a) and (c), error bars give the $95\%$ interval of the mean. Shaded areas: $95\%$ range of $200$ simulated corpora, each from a different draw of the posterior. Dashed line: closed-form prediction at the best fit; dotted lines: its $95\%$ posterior range.}
    \label{fig:model-vs-data-time-ev}
\end{figure*}

The model overestimates the average collaboration size (Fig.~\ref{fig:model-vs-data-time-ev}(c)): only $4$ of the $17$ yearly data points lie within the $95\%$ posterior-predictive band. The collaboration size is not a calibration target: the long-time size $q^*$ is fixed only indirectly, through the deletion and addition statistics, so its precision within the model (Sec.~\ref{sec:identifiability}) does not guarantee that it matches the data. We discuss possible sources of this discrepancy in Sec.~\ref{sec:discussion}. Moreover, the amount of authors in a scientific publication can depend on multiple factors such as the topic of research \cite{wuchty2007increasing, newman2001scientific} or existing funding \cite{ubfal2011impact}, rather than being a direct consequence of previous collaborations.

Nonetheless, the number of distinct authors in the data grows linearly at the rate predicted by the model, $a\approx1.40$ authors per paper at the best fit [$95\%$ credible interval $1.33$--$1.52$] (Fig.~\ref{fig:model-vs-data-time-ev}(b)). Since the final number of authors is one of the calibration targets, this agreement is partly built in; the calibration does not impose, however, that the data follow the predicted linear growth over the observation period as a whole. In detail, Fig.~\ref{fig:model-vs-data-time-ev}(b) shows how the data lie slightly below the model in the intermediate years and catch up in the last ones (inside the $95\%$ posterior-predictive band in $9$ of the $17$ years), when the number of authors added to modified teams increased.

Moreover, the average productivity of the data follows the predicted increase towards the long-time limit of Eq.~\eqref{eq:solution_k_t}, $\langle k\rangle_\infty\approx 3.3$ at the best fit (Fig.~\ref{fig:model-vs-data-time-ev}(c)), with $15$ of the $17$ yearly data points within the $95\%$ posterior-predictive band. The decrease of the observed $\langle k(t)\rangle$ in the last years mirrors the faster growth of $A(t)$ in the same years: the influx of new authors lowers the average productivity, consistently with the larger number of added authors in recent years.

\subsection{Collaboration size and productivity distributions}

\begin{figure*}[t]
    \centering
    \includegraphics[width=\textwidth]{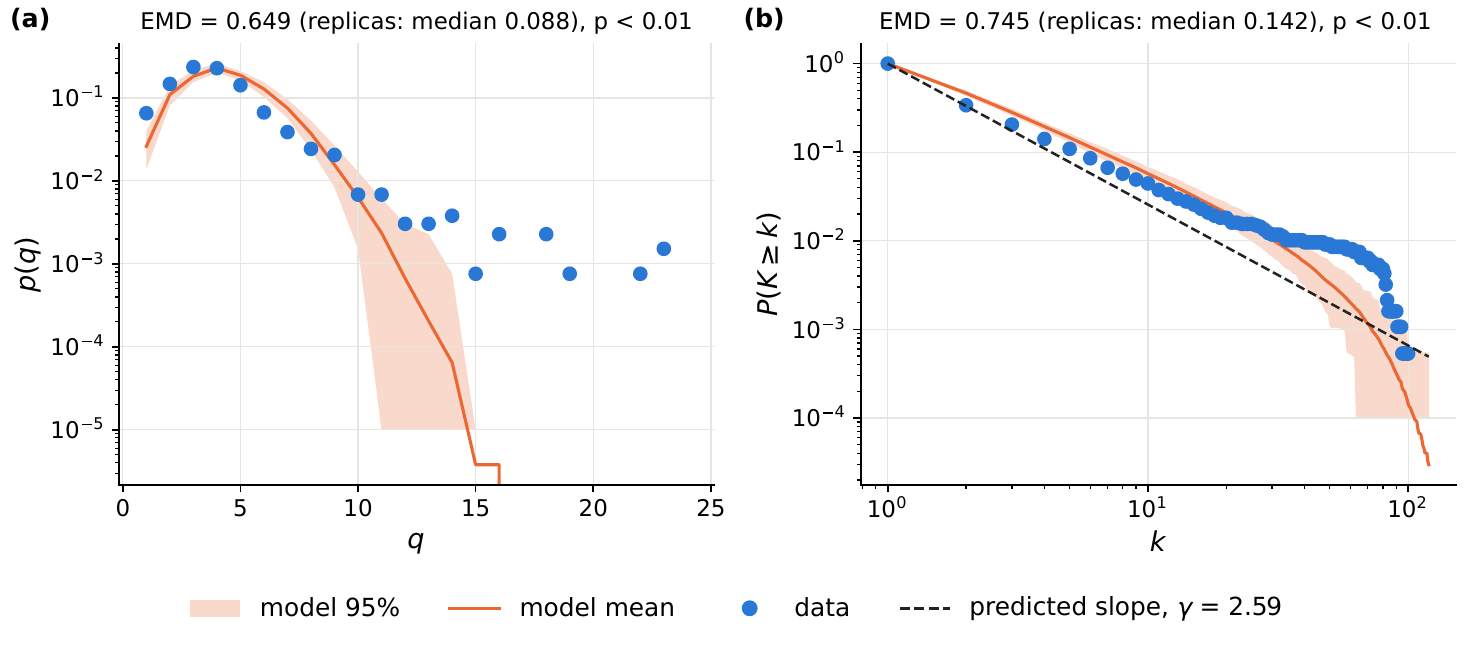}
    \caption{Distributions of (a) collaboration size and (b) productivity (complementary cumulative distribution) in the data and in $100$ simulations at the best fit (mean and $95\%$ range). The dashed line has the slope predicted by Eq.~\eqref{eq:gamma} at the best fit, $\gamma=2.59$. Panel titles give the Earth Mover's Distance between the data and the model mean, its median among model replicas, and the resulting $p$-value.}
    \label{fig:model-vs-data-distributions}
\end{figure*}

Figure~\ref{fig:model-vs-data-distributions} compares the observed and simulated distributions. According to the parametric bootstrap described above, both distributions are rejected ($p<0.01$). For the collaboration size [Fig.~\ref{fig:model-vs-data-distributions}(a)], the data contain $2.5$ times more single-author papers than the model ($6.5\%$ against $2.6\%$), as well as a tail of large collaborations ($3.2\%$ of papers with ten or more authors, against $0.9\%$) that Poisson additions cannot generate (see Sec.~\ref{sec:discussion}).

The data are, moreover, not in the stationary regime of Eq.~\eqref{eq:p_q-stationary}: at the best fit, with $\alpha=p_{\text{mod}}p_{\text{del}}\approx0.43$, the transient of Eq.~\eqref{eq:solution_q_t} has decayed only by a factor $(1317/59)^{-0.43}\approx0.26$ over the observation period. A larger corpus would be needed to compare the data with the stationary distribution itself.

For the productivity [Fig.~\ref{fig:model-vs-data-distributions}(b)], the data have more one-paper authors than the model ($66\%$ against $54\%$) and, at the same time, a more concentrated top: the $1\%$ most productive authors hold $24\%$ of the authorships in the data and $15\%$ in the model. The calibrated model thus captures the scale and the overall, average growth of the network, but not the heterogeneity of its distributions. We discuss possible causes of these discrepancies in Sec.~\ref{sec:discussion}.

\section{Discussion} \label{sec:discussion}

We have introduced a growth model for scientific collaboration networks in which a new paper starts from the team of authors of an existing one. The new paper is written either by an entirely new team, by the same team, or by a modified team in which some members leave and new or already active members join. The model is inspired by the duplication--divergence paradigm, mostly used to describe biological networks, and aims to represent the mechanisms that drive the formation of teams within a research institution.

We derived the stationary distribution of collaboration size, the tail of the productivity distribution, and the time evolution of the average collaboration size, the number of distinct authors and the average productivity. Uniform copying of papers acts as an effective preferential attachment on authors, which produces a power-law distribution of papers per author, the Lotka-like regularity long observed in bibliometric data~\cite{lotka1926frequency, simon1955class, price1976general}, without any explicit preference for productive authors. These expressions agree with numerical simulations, with small deviations in the tail of the productivity distribution. We also developed a simulation-based framework to infer the model parameters from empirical data, and applied it to the publication record of IFISC. The calibrated model describes the institute as a network of teams that evolve mostly by gradual change: about $78\%$ of the papers are modifications of earlier collaborations, about $19\%$ are written by exactly the same team as an earlier paper, and only a few percent are written by completely new teams. With these parameters, the model reproduces the growth of the number of distinct authors and of the average productivity, but it overestimates the average collaboration size and does not capture the full heterogeneity of the distributions of collaboration size and productivity.

The model has limitations that, as we discuss in SM, Sec.~\ref{sec:appendix_extensions}, are likely to contribute to these discrepancies. First, the number of added authors follows a Poisson distribution, which is narrower than the one observed in the data. A negative binomial distribution, with a broader tail, would leave $\langle q(t)\rangle$, $A(t)$, $\langle k(t)\rangle$ and $\gamma$ unchanged, since they depend on this distribution only through its mean $\lambda$; a preliminary test (SM, Sec.~\ref{sec:appendix_extensions_nb}) suggests that it could account for the excess of single-author papers and of large collaborations. Second, every past paper is equally likely to be copied, irrespective of its age, so that authors never leave the network and old collaborations remain as influential as recent ones. The model therefore describes the turnover of team members, but not the departure of researchers from the institution. An ageing function in the choice of the reference paper, as in Ref.~\cite{ramasco2004self}, would let researchers become inactive over time, and would agree with the markedly recent references of modified papers in the data. Third, existing authors could be re-added preferentially, according to their productivity or their recent activity, rather than uniformly, which could account for the concentrated top of the productivity distribution. (SM, Sec.~\ref{sec:appendix_extensions_reference}). Finally, applying the same framework to several research institutions would turn the model parameters into a quantitative description of how collaboration is organized in each of them, and would make it possible to explore how institutional policies, such as the recruitment of new groups or the stability of established teams, shape the resulting collaboration network.

Because the model omits these mechanisms, its parameters are effective rates: the values with which this minimal model reproduces the corpus, absorbing the effect of the missing mechanisms. The stiff combinations of parameters (Sec.~\ref{sec:identifiability}), which are tied to what the corpus records directly, such as the long-time collaboration size and the share of newcomers, must be reproduced by any model fitted to the same data, and are therefore expected to be robust to such omissions. The missing mechanisms are instead most likely to show up in the soft directions, namely in the individual values of $p_{\text{del}}$, $\lambda$ and $p_{\text{add,new}}$, which should be read as effective rates rather than as measured behaviour.

More generally, our results show that a few team-level mechanisms, calibrated on publication records, account for several network-level properties of a collaboration network. Our work connects the long tradition of growth models in network science with the organization of research within an institution. Combining such models with more detailed organizational data, and extending them along the lines discussed above, would help clarify how institutional organization defines the formation of scientific collaboration networks.

\begin{acknowledgments}
JGC acknowledges funding from Mar\'ia de Maeztu (MdM) Seal of Excellence (CEX2021-001164-M) via the FPI programme (Grant No. PRE2022-103304), funded by the MICIU/AEI/10.13039/501100011033. BAG acknowledges support from the Complex Human Behaviour Lab of Fondazione Bruno Kessler (FBK), Trento, Italy. JAML acknowledges support from the Modeling and Simulation of Socio-Technical Systems (MoST) unit of Fondazione Bruno Kessler (FBK), Trento, Italy.JFG acknowledges partial support from project CSxAI (PID2024-157526NB-I00) funded by MICIU/AEI/10.13039/501100011033/FEDER, UE.  
\end{acknowledgments}

\section*{Data availability}
The dataset and the code used to reproduce the results in this article can be found in the following Zenodo repository: \url{https://zenodo.org/records/24691937}

\bibliography{references-collabs.bib}

\clearpage
\onecolumngrid
\begin{center}
\textbf{\large Supplemental Material: Duplication--divergence-inspired model for growing scientific collaboration networks}
\end{center}
\vspace{1em}
\twocolumngrid

\setcounter{section}{0}
\setcounter{equation}{0}
\setcounter{figure}{0}
\setcounter{table}{0}
\renewcommand{\thesection}{S\arabic{section}}
\renewcommand{\theequation}{S\arabic{equation}}
\renewcommand{\thefigure}{S\arabic{figure}}
\renewcommand{\thetable}{S\arabic{table}}
\renewcommand{\theHsection}{S\arabic{section}}
\renewcommand{\theHequation}{S\arabic{equation}}
\renewcommand{\theHfigure}{S\arabic{figure}}
\renewcommand{\theHtable}{S\arabic{table}}

In this Supplemental Material, sections, equations, figures and tables numbered with the prefix S belong to this document; all other numbers refer to the main text.

\section{Derivation of the mean-field results}
\label{sec:expressions}

In this section we derive the results of Sec.~\ref{subsec:analytical} of the main text. We use the same notation, $\alpha = p_{\text{mod}}p_{\text{del}}$, $q^*=1+\lambda/p_{\text{del}}$, and identify time with the number of papers, $P(t)=t$. All quantities are ensemble averages and to lighten the notation we do not distinguish them from their values in a single realization.

\subsection{Average collaboration size, $\langle q(t)\rangle$}
\label{sec:appendix_average_authors_paper}

The size of the new paper depends on the branch of the dynamics. In the new-team and repetition branches it equals the size of the reference, whose average is $\langle q(t)\rangle$ because the reference is drawn uniformly at random. In the modification branch, the protected author and, on average, a fraction $1-p_{\text{del}}$ of the remaining $q-1$ authors are kept, and on average $\lambda$ authors are added. The expected size of the new paper is therefore
\begin{equation}
    \mathbb{E}[q_{\text{new}}] = \langle q(t)\rangle - \alpha\left(\langle q(t) \rangle -1\right) + p_{\text{mod}}\lambda .
    \label{eq:e_qnew}
\end{equation}
The total number of author--paper links, $S(t) = \sum_{\text{papers}} q = t\langle q(t)\rangle$, evolves as $S(t+1) = S(t) + \mathbb{E}[q_{\text{new}}]$, so that
\begin{equation}
    \langle q(t+1)\rangle - q^* = \left(1-\frac{\alpha}{t+1}\right)\left[\langle q(t)\rangle - q^*\right].
    \label{eq:recursion_q}
\end{equation}
This linear recursion can be solved exactly,
\begin{equation}
    \langle q(t)\rangle - q^* = \left[\langle q(t_0)\rangle - q^*\right]\frac{\Gamma(t_0+1)\,\Gamma(t+1-\alpha)}{\Gamma(t+1)\,\Gamma(t_0+1-\alpha)} ,
    \label{eq:exact_q}
\end{equation}
and for $t\gg 1$ the ratio of Gamma functions behaves as $(t/t_0)^{-\alpha}$, which recovers Eq.~\eqref{eq:solution_q_t}. Equivalently, in the continuum limit Eq.~\eqref{eq:recursion_q} becomes the rate equation~\eqref{eq:rate_q}. For $p_{\text{del}}=0$ the recursion reduces to $\langle q(t+1)\rangle = \langle q(t)\rangle + p_{\text{mod}}\lambda/(t+1)$, whose exact solution is
\begin{equation}
    \langle q(t)\rangle = \langle q(t_0)\rangle + p_{\text{mod}}\lambda\left(H_t - H_{t_0}\right),
    \label{eq-ap:solution-q_t-pdel-0}
\end{equation}
where $H_t$ is the $t$-th harmonic number. Since $H_t - H_{t_0}\simeq \log(t/t_0)$ only for $t_0\gg1$, the continuum approximation of Eq.~\eqref{eq:solution-q_t-pdel-0} overestimates $\langle q(t)\rangle$ by $p_{\text{mod}}\lambda(1-\gamma_{\text{E}})\approx 0.42\,p_{\text{mod}}\lambda$ when $t_0=1$, where $\gamma_{\text{E}}$ is the Euler--Mascheroni constant.

\subsection{Number of distinct authors, $A(t)$}
\label{sec:appendix_unique_authors}

New authors enter the network in two ways. With probability $p_{\text{new}}$ a new team is created, whose size is that of the reference, with average $\langle q(t)\rangle$. With probability $p_{\text{mod}}$ a modified team gains on average $\lambda$ authors, each of whom is new to the network with probability $p_{\text{add,new}}$. Hence
\begin{equation}
    \frac{dA}{dt} = p_{\text{new}} \langle q(t)\rangle + p_{\text{mod}}\,p_{\text{add,new}}\,\lambda .
\end{equation}
For $p_{\text{del}}>0$, integrating with Eq.~\eqref{eq:solution_q_t} gives Eq.~\eqref{eq:solution_A_t} (for $\alpha\neq1$), whose long-time behaviour is $A(t)\simeq a\,t$ with
\begin{equation}
    a = p_{\text{new}}\left(1+\frac{\lambda}{p_{\text{del}}}\right) + p_{\text{mod}}\,p_{\text{add,new}}\,\lambda .
    \label{eq:rate_a}
\end{equation}
For $p_{\text{del}}=0$, integrating with Eq.~\eqref{eq:solution-q_t-pdel-0} gives
\begin{equation}
\begin{aligned}
A(t) ={}& A(t_0) + \left[p_{\text{new}}\langle q(t_0)\rangle + p_{\text{mod}}p_{\text{add,new}}\lambda\right](t-t_0) \\
& + p_{\text{new}}p_{\text{mod}}\lambda \left[ t\log\left(\frac{t}{t_0}\right) -(t-t_0) \right],
\end{aligned}
\label{eq-ap:solution_A_t-pdel-0}
\end{equation}
so that $A(t)\simeq p_{\text{new}}p_{\text{mod}}\lambda\, t\log t$ for $p_{\text{new}}>0$, and $A(t)\simeq p_{\text{mod}}p_{\text{add,new}}\lambda\, t$ for $p_{\text{new}}=0$.

\subsection{Average productivity, $\langle k(t)\rangle$}
\label{sec:appendix_average_papers_author}

Counting the author--paper links from the side of the papers and from the side of the authors gives $S(t) = \sum_{\text{papers}} q = \sum_{\text{authors}} k$, i.e., $t\langle q(t)\rangle = A(t)\langle k(t)\rangle$, and therefore
\begin{equation}
    \langle k(t)\rangle = \frac{t\,\langle q(t)\rangle}{A(t)} .
    \label{eq-ap:k_t-total}
\end{equation}
For $p_{\text{del}}>0$, $\langle q(t)\rangle\to q^*$ and $A(t)/t\to a$, which gives Eq.~\eqref{eq:solution_k_t}. For $p_{\text{del}}=0$ and $p_{\text{new}}>0$, the leading terms are $\langle q(t)\rangle \simeq p_{\text{mod}}\lambda\log t$ and $A(t)\simeq p_{\text{new}}p_{\text{mod}}\lambda\, t\log t$, so that
\begin{equation}
    \lim_{t\to\infty}\langle k(t)\rangle = \frac{1}{p_{\text{new}}} ,
    \label{eq:solution_k_t_pdel_0}
\end{equation}
with corrections of order $1/\log t$. If $p_{\text{new}}=0$, instead, $\langle k(t)\rangle\simeq \langle q(t)\rangle/(p_{\text{mod}}p_{\text{add,new}}\lambda)$ grows logarithmically.

\subsection{Stationary distribution of collaboration size}
\label{sec:appendix_stationary_authors_papers}

Let $n(q,t) = t\,p(q,t)$ be the number of papers with $q$ authors. In the modification branch, a reference with $q'$ authors keeps the protected author plus a binomial number of the other $q'-1$ authors, so that the number of kept authors $s$ follows
\begin{equation}
    P_{\text{surv}}(s|q') = \binom{q'-1}{s-1}(1-p_{\text{del}})^{s-1}p_{\text{del}}^{q'-s},
\end{equation}
for $s=1,\dots,q'$,
and $m$ authors are added with Poisson probability $P_{\text{add}}(m) = e^{-\lambda}\lambda^{m}/m!$. The size of the new paper thus changes from $q'$ to $q$ with probability
\begin{equation}
    W(q'\to q) = \sum_{s=1}^{\min(q,q')}P_{\text{surv}}(s|q')\,P_{\text{add}}(q-s).
\end{equation}
Since the reference is drawn uniformly at random, the new paper has size $q$ with probability
\begin{equation}
    g(q,t) = (1-p_{\text{mod}})\,p(q,t) + p_{\text{mod}}\sum_{q'\geq 1} p(q',t)\, W(q'\to q),
\end{equation}
and the expected number of papers of size $q$ evolves as $\mathbb{E}[n(q,t+1)\,|\,n(\cdot,t)] = n(q,t) + g(q,t)$. Writing $n(q,t)=t\,p(q,t)$ gives
\begin{equation}
\begin{aligned}
    (t+1)&\left[p(q,t+1)-p(q,t)\right] \\
    &= p_{\text{mod}}\left[\sum_{q'\geq 1} p(q',t)\,W(q'\to q)-p(q,t)\right],
\end{aligned}
\end{equation}
so that a stationary distribution must satisfy
\begin{equation}
    p_{\text{st}}(q) = \sum_{q'\geq 1} p_{\text{st}}(q')\,W(q'\to q).
    \label{eq:stationarity}
\end{equation}

To solve Eq.~\eqref{eq:stationarity} we introduce the generating function of $q-1$, $\psi(z) = \sum_{q\geq 1}p_{\text{st}}(q)\,z^{q-1}$. The number of kept non-protected authors is a binomial thinning of $q'-1$ with retention probability $1-p_{\text{del}}$, whose generating function is obtained by the substitution $z\to f(z) \coloneqq 1-(1-p_{\text{del}})(1-z)$, and the Poisson additions multiply the generating function by $e^{\lambda(z-1)}$. Equation~\eqref{eq:stationarity} is therefore equivalent to
\begin{equation}
    \psi(z) = e^{\lambda(z-1)}\,\psi\big(f(z)\big).
\end{equation}
Iterating this relation $n$ times, and using $f^{(j)}(z)-1 = (1-p_{\text{del}})^{j}(z-1)$ for the $j$-th iterate of $f$, gives
\begin{equation}
    \psi(z) = \exp\!\left[\lambda(z-1)\sum_{j=0}^{n-1}(1-p_{\text{del}})^{j}\right]\psi\big(f^{(n)}(z)\big).
\end{equation}
For $p_{\text{del}}>0$, $f^{(n)}(z)\to1$ as $n\to\infty$ and $\psi(1)=1$, so that the stationary solution is unique and equal to
\begin{equation}
    \psi(z) = \exp\!\left[\frac{\lambda}{p_{\text{del}}}(z-1)\right],
\end{equation}
the generating function of a Poisson distribution of mean $\Lambda=\lambda/p_{\text{del}}$. This proves Eq.~\eqref{eq:p_q-stationary}. The same argument applies to any distribution of added authors with generating function $G(z)$, for which $\psi(z)=\prod_{j\geq0}G\big(f^{(j)}(z)\big)$.

\subsection{Tail of the productivity distribution}
\label{sec:appendix_tail_papers_authors}

Consider an author $i$ with productivity $k_i$ at time $t$. The reference paper contains $i$ with probability $k_i/t$. In the repetition branch, $i$ is then an author of the new paper. In the modification branch, $i$ is kept with probability $1/q + (1-1/q)(1-p_{\text{del}}) = 1 - p_{\text{del}}(1-1/q)$, where $q$ is the size of the reference and $1/q$ the probability that $i$ is the protected author. In addition, $i$ can be added to a modified team as an existing author, with probability $p_{\text{mod}}\lambda(1-p_{\text{add,new}})/A(t)$. Altogether,
\begin{equation}
\begin{aligned}
\mathbb{E}&\left[k_i(t+1)-k_i(t)\right] \\
={}& \frac{1}{t}\sum_{j\ni i}\left\{p_{\text{rep}} + p_{\text{mod}}\left[1-p_{\text{del}}\left(1-\frac{1}{q_j}\right)\right]\right\} \\
& + \frac{p_{\text{mod}}\lambda(1-p_{\text{add,new}})}{A(t)},
\end{aligned}
\end{equation}
where the sum runs over the $k_i$ papers of author $i$. One can check that summing this expression over all authors and adding the contribution of newcomers recovers Eq.~\eqref{eq:e_qnew}. The papers of an author are a size-biased sample of all papers, so that, in a mean-field approximation, $\sum_{j\ni i} 1/q_j \simeq k_i/\langle q\rangle$; on average this identity is exact, since for a size-biased sample $\mathbb{E}_{\text{sb}}[1/q]=\langle q\cdot q^{-1}\rangle/\langle q\rangle=1/\langle q\rangle$ whatever the distribution of $q$, so that $B$, and hence $\gamma$, depends on the distribution of added authors only through its mean. For long times, with $\langle q\rangle\to q^*$ and $A(t)\simeq a\,t$, we obtain
\begin{equation}
    \frac{dk_i}{dt} = B\,\frac{k_i}{t} + \frac{c}{t},
\end{equation}
with $B$ given by Eq.~\eqref{eq:B} and $c = p_{\text{mod}}\lambda(1-p_{\text{add,new}})/a$.
For an author who enters the network at time $s$ with $k_i(s)=k_0$, the solution is
\begin{equation}
    k_i(t) = \left(k_0+\frac{c}{B}\right)\left(\frac{t}{s}\right)^{B}-\frac{c}{B},
\end{equation}
so that an author reaches productivity $k$ if the author entered before $s = t\,\big[(k+c/B)/k_0'\big]^{-1/B}$, with $k_0'=k_0+c/B$. Since authors enter the network at a constant rate, entry times are uniformly distributed in $[0,t]$, and
\begin{equation}
    \Pr(K\geq k) = \left(\frac{k+c/B}{k_0'}\right)^{-1/B} \sim k^{-1/B} .
\end{equation}
The tail is thus a power law shifted by $c/B$, which approaches a pure power law only for $k\gg c/B$; this curvature contributes to making the exponents fitted at moderate $k$ smaller than $\gamma$ (main text, Sec.~\ref{subsec:validation}).
Differentiating with respect to $k$ gives $p(k)\sim k^{-\gamma}$ with $\gamma = 1+1/B$, and using $1-1/q^* = \lambda/(\lambda+p_{\text{del}})$ yields Eq.~\eqref{eq:gamma}. Note that $B>0$ whenever $p_{\text{new}}<1$, because the protected author always provides a contribution proportional to $k_i$; for $p_{\text{new}}=1$ every author writes a single paper.

\section{Convergence of $\langle q(t) \rangle$}
\label{sec:appendix_convergence_authors_paper}

According to Eq.~\eqref{eq:solution_q_t}, $\langle q(t)\rangle$ is within a tolerance $\beta$ of its stationary value $q^*$ when
\begin{equation}
    \left|\langle q(t_0)\rangle - q^* \right|\left(\frac{t}{t_0}\right)^{-p_{\text{mod}}p_{\text{del}}} \leq \beta .
\end{equation}
For simulations of $t_{\max}$ papers, convergence within the simulated time therefore requires
\begin{equation}
    p_{\text{mod}}\,p_{\text{del}} \geq \frac{\log\left(\left|\langle q(t_0) \rangle - q^*\right|/\beta\right)}{\log(t_{\max}/t_0)} ,
\end{equation}
which is a condition we used to choose the parameter sets of Table~\ref{tab:parameters-2} of the main text for the case of high and medium values of $p_{\text{mod}}$.

\section{Numerical validation for $p_{\text{del}} = 0$}
\label{sec:appendix_verification_del_0}

We validate here the results for $p_{\text{del}}=0$ with $100$ simulations of $20\,000$ papers, for $p_{\text{new}} = 0.2$, $p_{\text{rep}} = 0.1$, $p_{\text{add,new}} = 0.5$ and $\lambda = 2.5$, starting from a single paper with $5$ authors.

\begin{figure*}[tp]
    \centering
    \includegraphics[width=\textwidth]{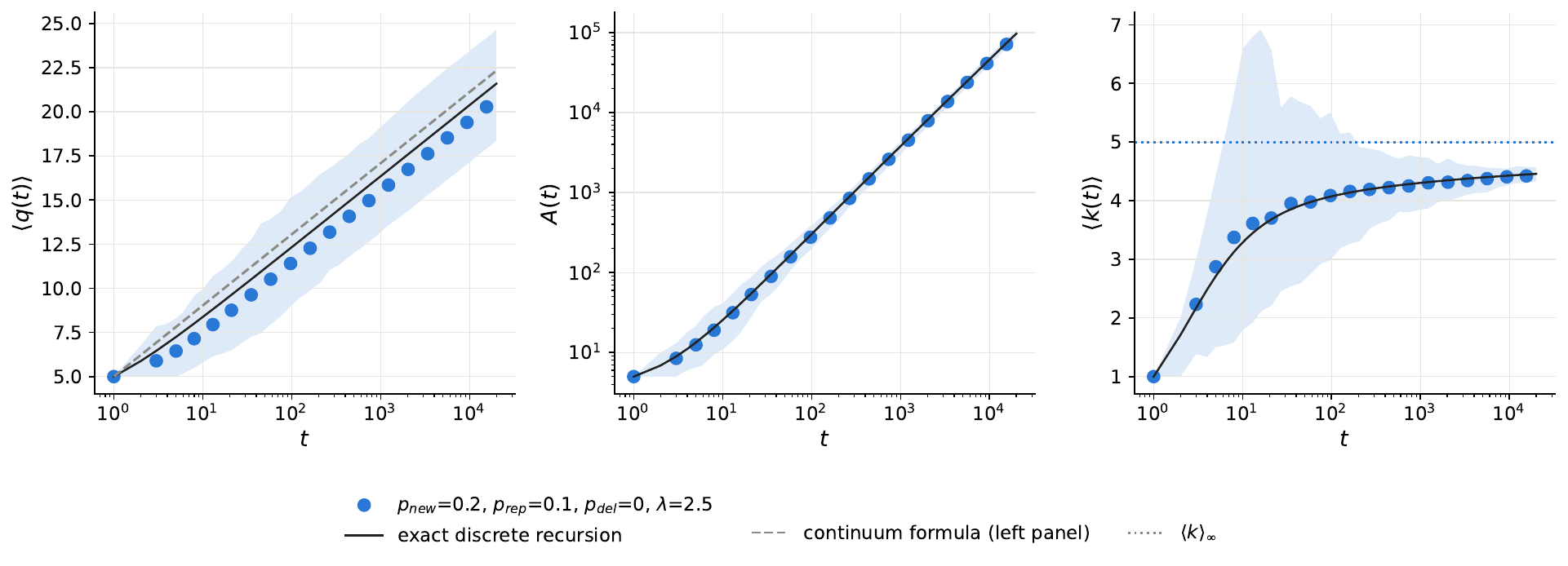}
    \caption{Time evolution for $p_{\text{del}}=0$ of the average collaboration size (left), the number of distinct authors (middle) and the average productivity (right), with $p_{\text{new}}=0.2$, $p_{\text{rep}}=0.1$, $p_{\text{add,new}}=0.5$ and $\lambda=2.5$, from a single paper with $5$ authors. Markers: mean of $100$ simulations; shaded band: range of $95\%$ of the simulations. Solid line: exact solution of the discrete mean-field recursion. Dashed line: continuum formula, Eq.~\eqref{eq:solution-q_t-pdel-0}. Dotted line: limit $1/p_{\text{new}}$ of the average productivity.}
    \label{fig:pdel_0}
\end{figure*}

Figure~\ref{fig:pdel_0} compares the simulations with the exact solution of the discrete mean-field recursion of Sec.~\ref{sec:expressions} and with the continuum formulas of the main text. The average collaboration size grows logarithmically, as predicted. The continuum formula, Eq.~\eqref{eq:solution-q_t-pdel-0}, lies above the exact solution, Eq.~\eqref{eq-ap:solution-q_t-pdel-0}, by a constant offset, the difference between $\log t$ and the harmonic number: for $t_0=1$ it equals $(1-\gamma_{\text{E}})p_{\text{mod}}\lambda\approx 0.74$. The simulations lie a further $0.87$ below the exact solution at $t=20\,000$ ($20.72$ against $21.59$). This gap comes from a detail of the simulation that the mean-field equations ignore: when an existing author is to be added to a modified team but every author of the network is already in it, which happens in the first steps, the addition is skipped (main text, Sec.~\ref{sec:model}). With $p_{\text{add,new}}=1$, so that no existing author is ever needed, the gap shrinks to $0.13$ ($21.46$ against $21.59$).

The number of distinct authors follows the exact solution, including the $t\log t$ growth caused by the increasing size of new teams, to within $4\%$ ($93\,400$ against $96\,900$ at $t=20\,000$), and the average productivity, Eq.~\eqref{eq-ap:k_t-total}, to within $1\%$ ($4.43$ against $4.46$). At $t=20\,000$ the average productivity is still below its limit $1/p_{\text{new}}=5$, Eq.~\eqref{eq:solution_k_t_pdel_0}, which is approached only with corrections of order $1/\log t$.

\section{Empirical support for the model assumptions}
\label{sec:appendix_justification}

\begin{figure*}[tp]
    \centering
    \includegraphics[width=\textwidth]{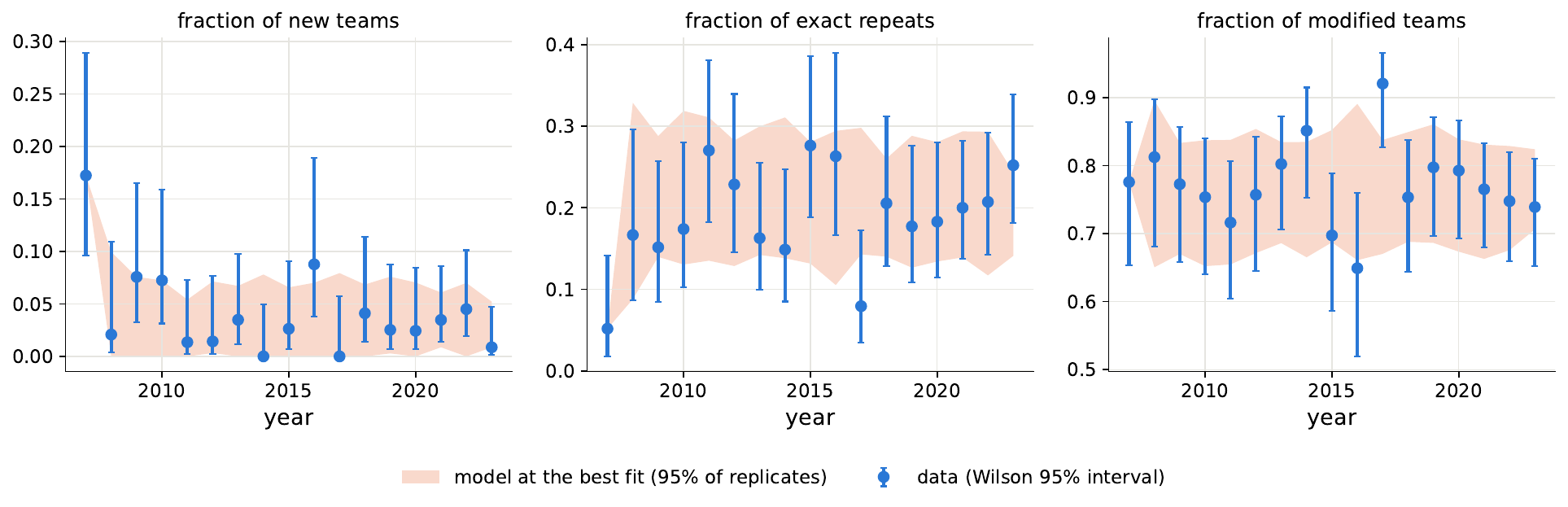}
    \caption{Time evolution of the fractions of new, repeated and modified papers in the dataset as well as with the model at best fit.}
    \label{fig:fractions-over-time}
\end{figure*}

The model assumes that the probabilities of its three branches are constant in time. Figure~\ref{fig:fractions-over-time} shows the time evolution of the fractions of new, repeated and modified papers in the data as the corpus grows. Despite some variations in specific years, the fractions remain roughly constant, which supports the assumption of constant probabilities. The number of authors added to modified papers, instead, increased in the last years (Sec.~\ref{sec:appendix_extensions_time}).

\begin{figure*}[tbp]
    \centering
    \includegraphics[width=\textwidth]{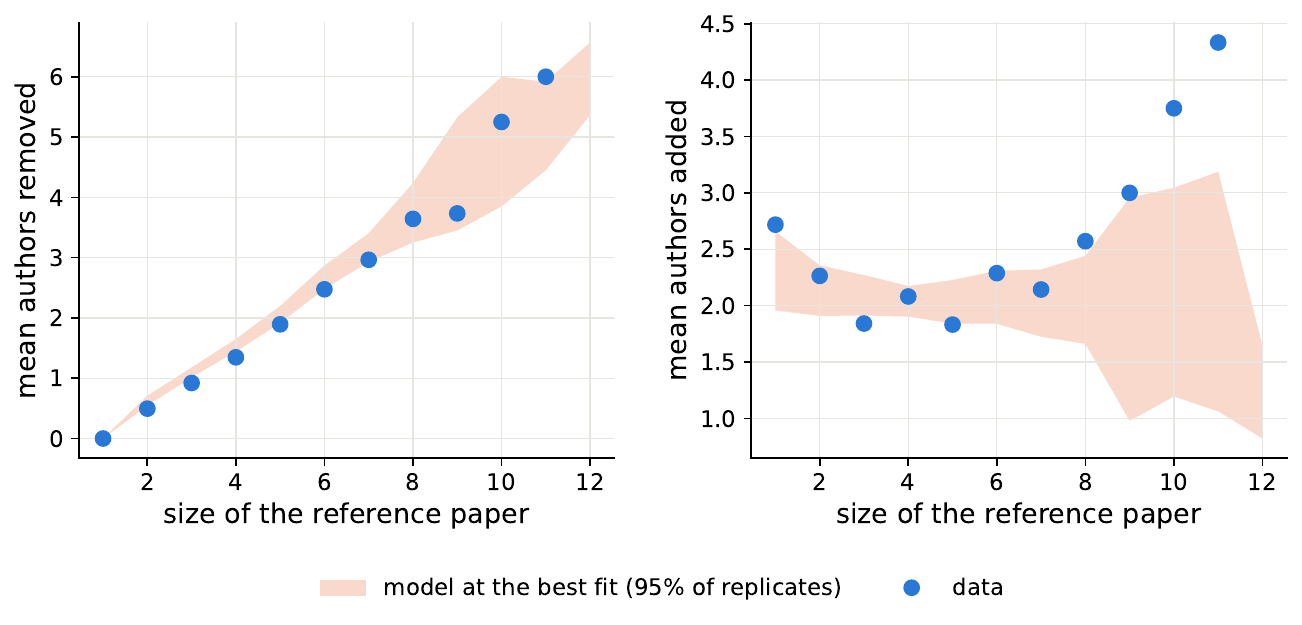}
    \caption{Average number of removed (left) and added (right) authors per modified paper, as a function of the size of the reference paper, in the data and in $50$ simulations at the best fit ($95\%$ range). The reference of each modified paper is estimated as the earlier paper with the largest overlap of authors.}
    \label{fig:removals-and-additions-vs-degree}
\end{figure*}

The model also assumes that, in the modification branch, each author of the reference is removed independently, so that the number of removed authors grows with the size of the reference, whereas the number of added authors does not depend on it. Figure~\ref{fig:removals-and-additions-vs-degree} supports both assumptions: in the data, the average number of removed authors grows linearly with the size of the reference, as in simulations of the calibrated model, whereas the average number of added authors is approximately independent of it, at least up to sizes of about eight authors, beyond which it increases and the statistics are poor (see also Sec.~\ref{sec:appendix_extensions_authors}).

\section{Details of the calibration}
\label{sec:appendix_calibration}

This section gives the numerical settings of the calibration of Sec.~\ref{sec:calibration} of the main text and the analyses that support Sec.~\ref{sec:identifiability}. Throughout, $n_{\text{rep}}$ denotes the number of independent simulations, with different random seeds, over which a quantity is averaged. Every simulated corpus is seeded with the $59$ papers of the initial year (2007) and grown to $1317$ papers, the size of the observed corpus.

\subsection{Distance and global search}
\label{sec:appendix_distance}
\label{sec:appendix_global}

The scale $\sigma_i$ of each term of Eq.~\eqref{eq:fitness} is the standard deviation of the statistic $s_i$ over $50$ simulations at the best fit of a preliminary global search: $\sigma_{s_A}=50.8$, $\sigma_{s_{\text{new}}}=0.0057$, $\sigma_{s_{\text{rep}}}=0.0119$, $\sigma_{s_{\text{del}}}=0.0091$, $\sigma_{s_{\text{add}}}=0.059$ and $\sigma_{s_{\text{add,new}}}=0.0093$. The global search evaluates the distance on $1000$ points of a Latin hypercube design~\cite{mckay1979comparison} with $n_{\text{rep}}=20$, and refines the best $10$ with the Nelder--Mead algorithm~\cite{nelder1965simplex} (at most $500$ iterations, tolerances $10^{-3}$ in parameter space and $10^{-4}$ in the objective).

\subsection{Approximate Bayesian computation}
\label{sec:appendix_abc}
\label{sec:appendix_posterior}

\begin{figure*}[tp]
    \centering
    \includegraphics[width=\textwidth]{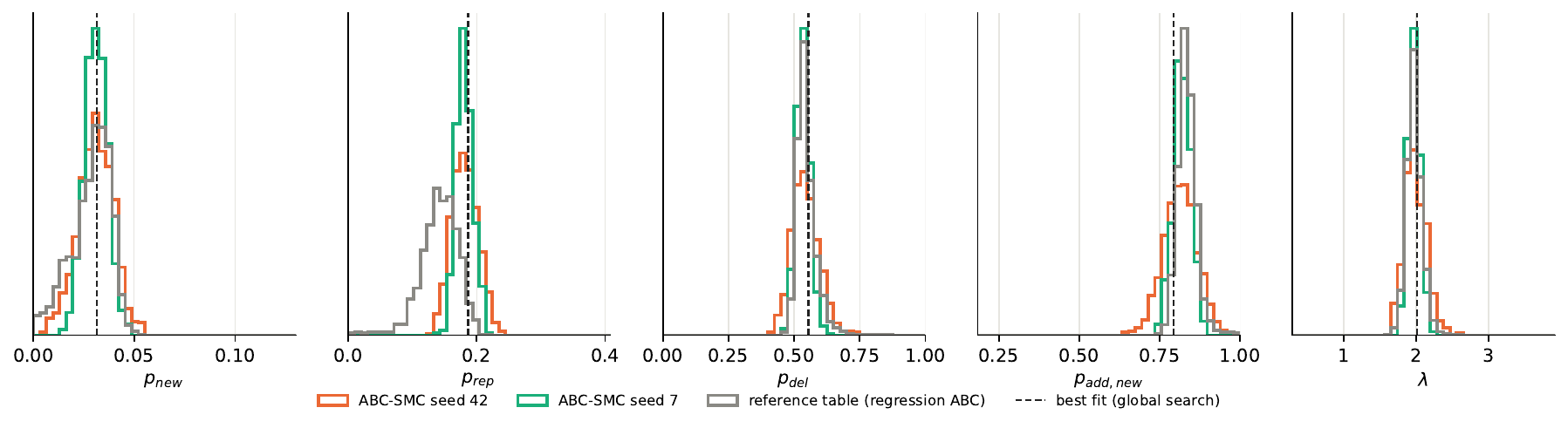}
    \caption{Marginal posterior distributions of the five parameters from the two ABC-SMC runs with different random seeds, and from regression-adjusted ABC on a reference table of prior simulations (Sec.~\ref{sec:appendix_recovery}). Dashed line: best fit of the global search.}
    \label{fig:posterior}
\end{figure*}

The ABC-SMC scheme~\cite{beaumont2002approximate, toni2008approximate} starts from a uniform prior over the parameter box and propagates $1000$ particles through $20$ generations. In each generation, the tolerance $\varepsilon$ is the weighted median of the distances of the current population; new particles are proposed with a Gaussian kernel whose covariance is twice the weighted covariance of the current population, proposals outside the prior support are rejected, and a proposal is accepted if its distance, evaluated with $n_{\text{rep}}=3$ new simulations, is below $\varepsilon$. The importance weight of particle $i$ in generation $g$ is
\begin{equation}
\omega_i^{(g)} \propto \Big[\textstyle\sum_j \omega_j^{(g-1)}\,\mathcal{K}\big(\Theta_i^{(g)}-\Theta_j^{(g-1)}\big)\Big]^{-1},
\label{eq:weights_abc}
\end{equation}
with $\mathcal{K}$ the perturbation kernel. The run uses about $4.2\times10^{5}$ simulations. Figure~\ref{fig:posterior} shows the marginal posteriors.

Because the simulator is stochastic, the distance between data and simulations at the best fit is not zero: its median is $5.4$ and its $95\%$ range $[3.6, 7.9]$. The final tolerance, $\varepsilon=8.4$, lies just above this noise floor, below which the tolerance would no longer discriminate between parameter values. A second run with a different random seed, continued to $\varepsilon=5.5$, gives the same medians to within $0.005$ and intervals $1.6$--$1.9$ times narrower (Table~\ref{tab:posteriors}). The main text reports the first run, which is the conservative choice.

\begin{table}[tbp]
\caption{Posterior medians and $95\%$ credible intervals of two ABC-SMC runs with different random seeds. The first is reported in the main text.}
\label{tab:posteriors}
\begin{ruledtabular}
\begin{tabular}{lccc}
 & Median & $95\%$ CI, run 1 & $95\%$ CI, run 2 \\
 & & ($\varepsilon=8.4$) & ($\varepsilon=5.5$) \\
\colrule
$p_{\text{new}}$      & 0.031 & [0.013, 0.048] & [0.020, 0.042] \\
$p_{\text{rep}}$      & 0.181 & [0.143, 0.223] & [0.158, 0.209] \\
$p_{\text{del}}$      & 0.539 & [0.448, 0.652] & [0.484, 0.590] \\
$\lambda$             & 1.98  & [1.72, 2.32]   & [1.82, 2.14] \\
$p_{\text{add,new}}$  & 0.814 & [0.709, 0.913] & [0.763, 0.874] \\
\end{tabular}
\end{ruledtabular}
\end{table}

\begin{figure*}[tp]
    \centering
    \includegraphics[width=\textwidth]{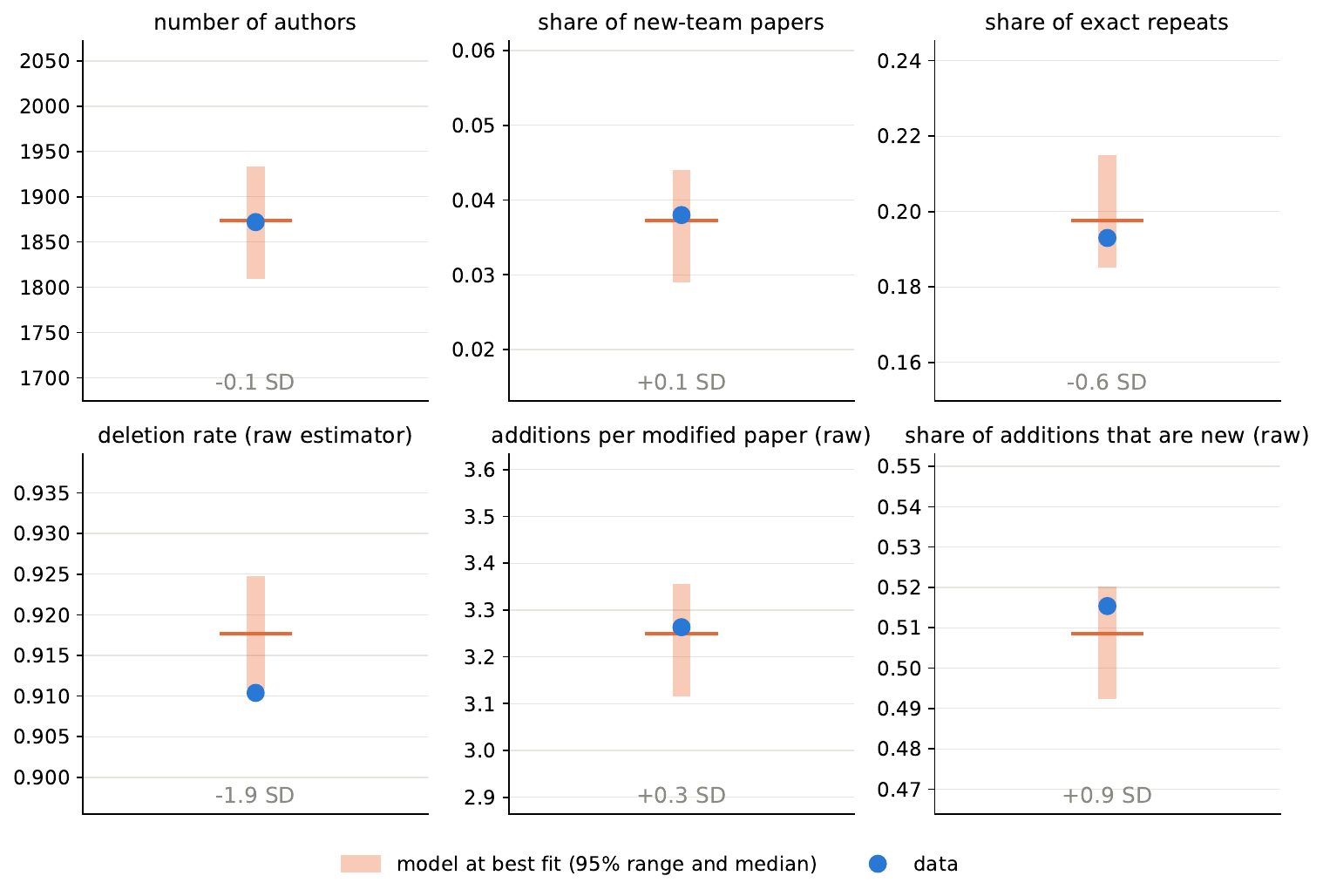}
    \caption{The six calibration statistics at the best fit. Bars: $95\%$ range and median over $30$ simulations; dots: data. Numbers give $(s^{\text{obs}}-\bar s)/\mathrm{SD}(s)$.}
    \label{fig:best_fit_summaries}
\end{figure*}

\textit{Fit of the calibration targets.} All six statistics of the data lie within two standard deviations of their simulated mean at the best fit (Fig.~\ref{fig:best_fit_summaries}). The largest deviation is that of the deletion statistic, at $-1.9$ standard deviations, just below the $95\%$ range of the simulations.

\subsection{Bias of the summary statistics}
\label{sec:appendix_bias}

\begin{figure*}[tp]
    \centering
    \includegraphics[width=\textwidth]{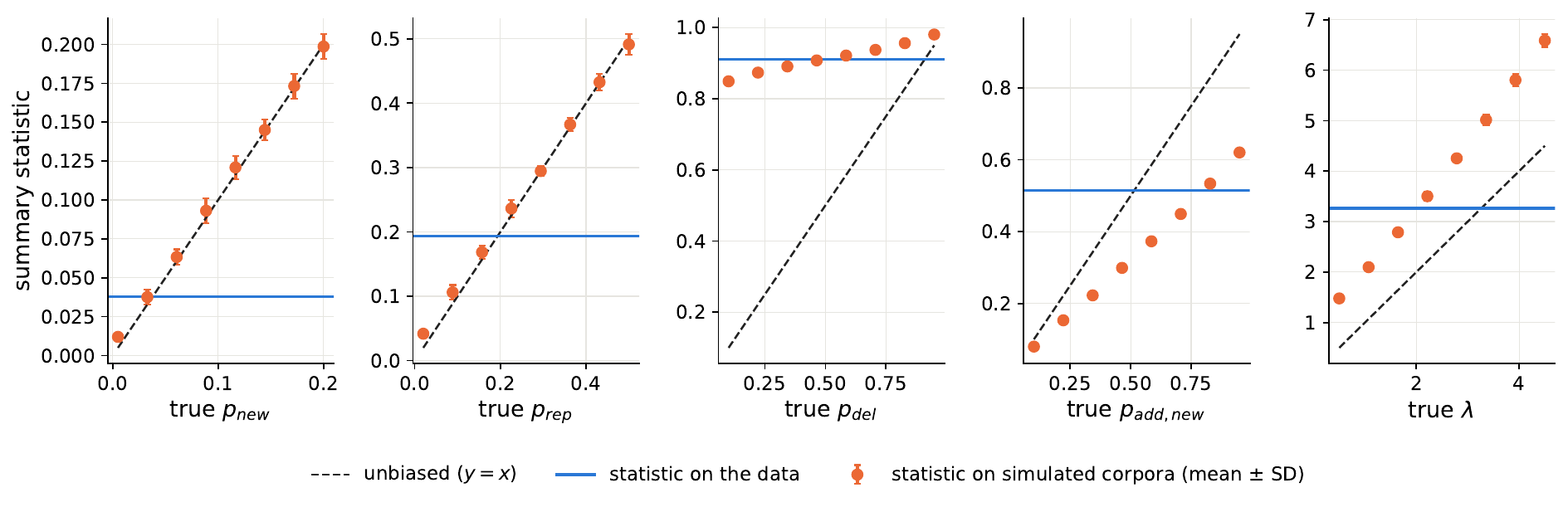}
    \caption{Summary statistics on simulated corpora (mean $\pm$ standard deviation over $20$ corpora) as a function of the true value of the corresponding parameter, with the other parameters at the best fit. Dashed diagonal: an unbiased estimator ($y=x$ line). Horizontal line: the value in the data.}
    \label{fig:estimator_bias}
\end{figure*}

To show why the statistics of Sec.~\ref{sec:estimators} are not parameter estimates, we vary one parameter at a time, with the others at the best fit, and compute the statistics on $20$ simulated corpora per value (Fig.~\ref{fig:estimator_bias}). The statistics $s_{\text{new}}$ and $s_{\text{rep}}$ follow $p_{\text{new}}$ and $p_{\text{rep}}$ closely. The other three do not: $s_{\text{add}}$ overestimates $\lambda$, $s_{\text{add,new}}$ underestimates $p_{\text{add,new}}$, and $s_{\text{del}}$ barely responds to $p_{\text{del}}$, because without knowledge of the true reference, deletions are counted against every earlier paper that overlaps with the modified one. The calibration corrects these biases jointly for all parameters.

\subsection{Profiles of the objective function}
\label{sec:appendix_multimodality}

\begin{figure*}[tp]
    \centering
    \includegraphics[width=\textwidth]{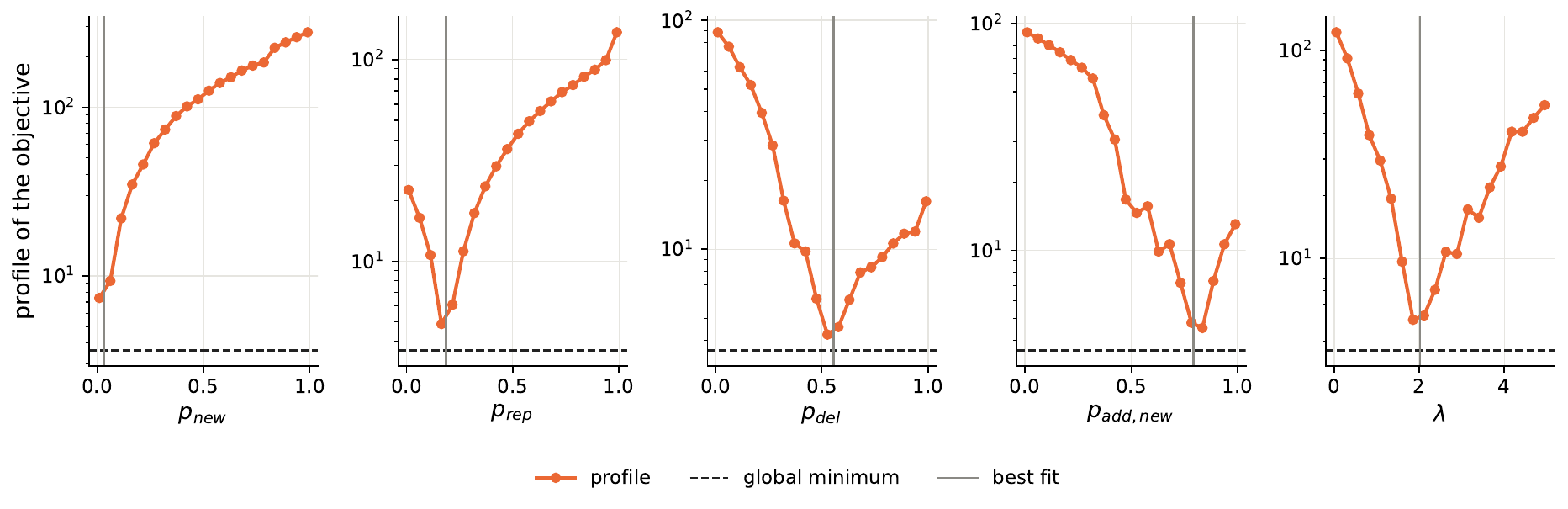}
    \caption{Profiles of the objective function: minimum distance attainable when one parameter is fixed and the other four are re-optimized (logarithmic scale). Horizontal dashed line: global minimum; vertical line: best fit.}
    \label{fig:profiles}
\end{figure*}

For each parameter, we fix it on a grid of $20$ values spanning the central $98\%$ of its range and re-optimize the other four with the Nelder--Mead algorithm ($200$ iterations, tolerance $10^{-3}$, $n_{\text{rep}}=30$). This is a profile of the distance of Eq.~\eqref{eq:fitness}, not a profile likelihood. Every profile has a single, well-defined minimum at the best fit (Fig.~\ref{fig:profiles}). The minima of the profiles lie above the global minimum (dashed line in Fig.~\ref{fig:profiles}) because the profiles are evaluated on a regular grid of parameter values, which in general does not contain the best fit, whereas the global minimum was obtained with a more precise optimization.

\subsection{Fisher information and sloppiness}
\label{sec:appendix_sloppiness}

Approximating the six summary statistics as Gaussian with mean $\bar s(\Theta)$ and covariance $\Sigma$~\cite{wood2010statistical}, with $\Sigma$ taken as independent of $\Theta$, the Fisher information with respect to the logarithms of the parameters is
\begin{equation}
F = J^\top \Sigma^{-1} J, \qquad J_{ij} = \frac{\partial \bar s_i}{\partial \log\Theta_j}.
\end{equation}
The Jacobian $J$ measures how much each statistic moves when a parameter is changed, and $\Sigma^{-1}$ expresses these changes in units of the simulator's noise; $F$ is thus the Gauss--Newton approximation of the curvature of the negative log-likelihood, i.e., of half the $\chi^2$ misfit, at the best fit~\cite{Gutenkunst2007, transtrum2015perspective}. We use it rather than the curvature of the distance of Eq.~\eqref{eq:fitness}, which is not differentiable at zero residuals. We estimate $J$ by central finite differences with a step of $5\%$ of the prior range of each parameter, $n_{\text{rep}}=200$ and common random numbers on the two sides of each difference, and $\Sigma$ from $200$ simulations at the best fit; a step of $2\%$ gives the same condition number.

\begin{figure*}[tp]
    \centering
    \includegraphics[width=\textwidth]{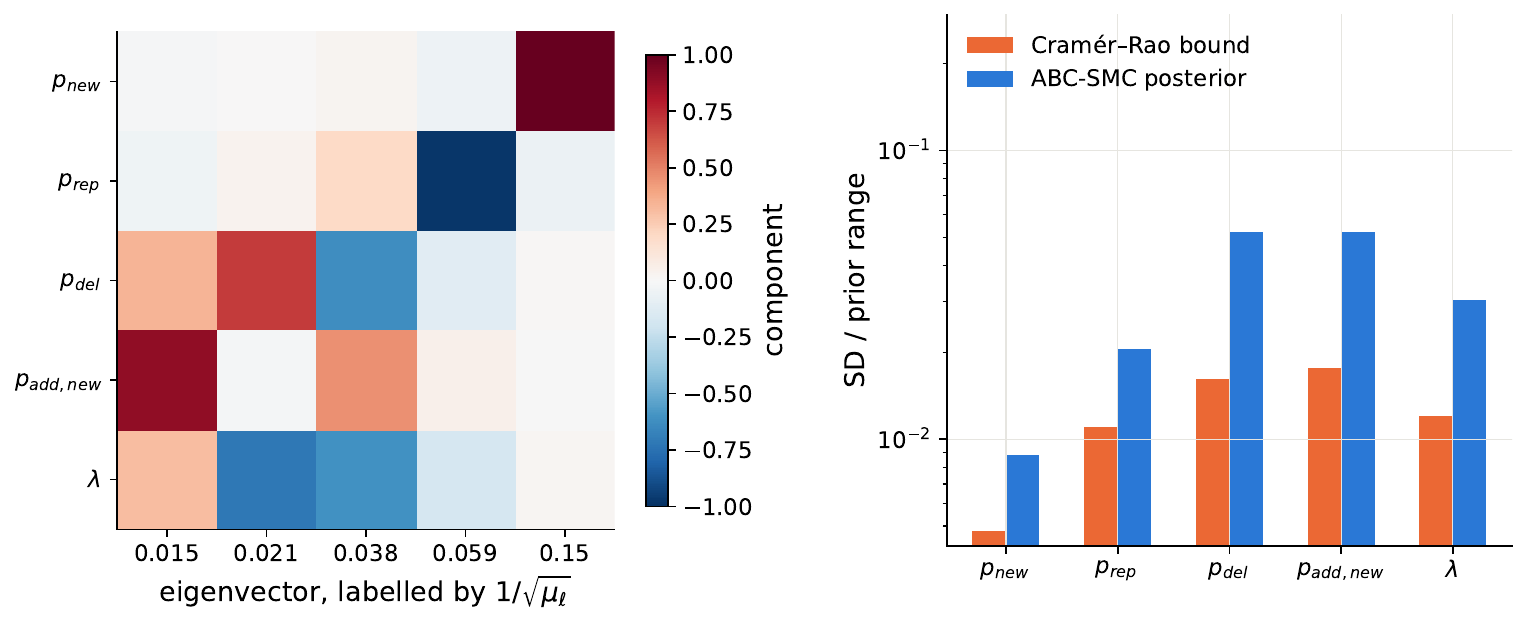}
    \caption{Fisher information at the best fit. Left: eigenvectors for the six calibration statistics, labelled by $1/\sqrt{\mu_\ell}$. Right: standard deviation of each parameter, as a fraction of its prior range, from the Cram\'er--Rao bound and from the ABC-SMC posterior.}
    \label{fig:sloppiness}
\end{figure*}

Table~\ref{tab:fisher} lists the eigenvalues and normalized eigenvectors of $F$. Since $F$ is computed with respect to the logarithms of the parameters, a change $\delta$ in $\log\Theta_j$ multiplies $\Theta_j$ by $e^{\delta}\simeq1+\delta$, so the half-lengths $1/\sqrt{\mu_\ell}$ are relative uncertainties. The second eigenvector, $v_2\simeq(\log p_{\text{del}}-\log\lambda)/\sqrt2$, has equal and opposite components on the two parameters, so a step of one half-length, $0.021$, along it changes $p_{\text{del}}$ and $\lambda$ by $0.021/\sqrt2\simeq1.5\%$ in opposite directions, and their ratio $\lambda/p_{\text{del}}$ by $3\%$. On the other axes the components on $p_{\text{del}}$ and $\lambda$ are nearly equal, so moving along them leaves the ratio almost unchanged, and $3\%$ is the minimal uncertainty on $\lambda/p_{\text{del}}$. Since $q^*=1+\lambda/p_{\text{del}}$ and the constant does not vary, the corresponding absolute uncertainty, $0.03\times3.63=0.11$, carries over to $q^*=4.63$, a relative uncertainty of about $2\%$. Figure~\ref{fig:sloppiness} shows the eigenvectors and the standard deviation of each parameter from the Cram\'er--Rao bound and from the ABC-SMC posterior.

In the ABC-SMC posterior, $p_{\text{del}}$ and $\lambda$ increase together and $p_{\text{add,new}}$ decreases (Table~\ref{tab:correlations}), along the third eigenvector of Table~\ref{tab:fisher}. Of the derived quantities, $q^*$ and $\langle k\rangle_\infty$ depend mainly on the stiff combinations and have relative posterior spreads of $3.5\%$ and $3.7\%$. The exponent $\gamma$ depends, through Eq.~\eqref{eq:gamma}, also on the common scale of $\lambda$ and $p_{\text{del}}$, but this term is only part of $B$, and its relative spread is $2.9\%$.

\begin{table}[tbp]
\caption{Eigenvalues $\mu_\ell$ and normalized eigenvectors of the Fisher information in log-parameter space at the best fit, with the half-length $1/\sqrt{\mu_\ell}$ of the corresponding axis of the one-standard-deviation ellipsoid.}
\label{tab:fisher}
\begin{ruledtabular}
\begin{tabular}{cccrrrrr}
$\ell$ & $\mu_\ell$ & $1/\sqrt{\mu_\ell}$ & $p_{\text{new}}$ & $p_{\text{rep}}$ & $p_{\text{del}}$ & $p_{\text{add,new}}$ & $\lambda$ \\
\colrule
1 & 4363 & 0.015 & $-0.01$ & $-0.04$ & $0.34$ & $0.89$ & $0.31$ \\
2 & 2235 & 0.021 & $0.01$ & $0.04$ & $0.70$ & $-0.02$ & $-0.71$ \\
3 & 681 & 0.038 & $0.03$ & $0.20$ & $-0.62$ & $0.46$ & $-0.61$ \\
4 & 284 & 0.059 & $-0.05$ & $-0.98$ & $-0.11$ & $0.05$ & $-0.17$ \\
5 & 42.9 & 0.153 & $1.00$ & $-0.06$ & $0.01$ & $0.00$ & $0.02$ \\
\end{tabular}
\end{ruledtabular}
\end{table}

\begin{table}[tbp]
\caption{Correlation coefficients between parameters in the ABC-SMC posterior of the main text.}
\label{tab:correlations}
\begin{ruledtabular}
\begin{tabular}{lccccc}
 & $p_{\text{new}}$ & $p_{\text{rep}}$ & $p_{\text{del}}$ & $p_{\text{add,new}}$ & $\lambda$ \\
\colrule
$p_{\text{new}}$     & 1 & 0.18 & 0.14 & $-0.23$ & 0.13 \\
$p_{\text{rep}}$     &   & 1    & 0.32 & $-0.26$ & 0.40 \\
$p_{\text{del}}$     &   &      & 1    & $-0.91$ & 0.90 \\
$p_{\text{add,new}}$ &   &      &      & 1       & $-0.87$ \\
$\lambda$            &   &      &      &         & 1 \\
\end{tabular}
\end{ruledtabular}
\end{table}

\subsection{Recovery of known parameters}
\label{sec:appendix_recovery}

\begin{figure*}[tp]
    \centering
    \includegraphics[width=\textwidth]{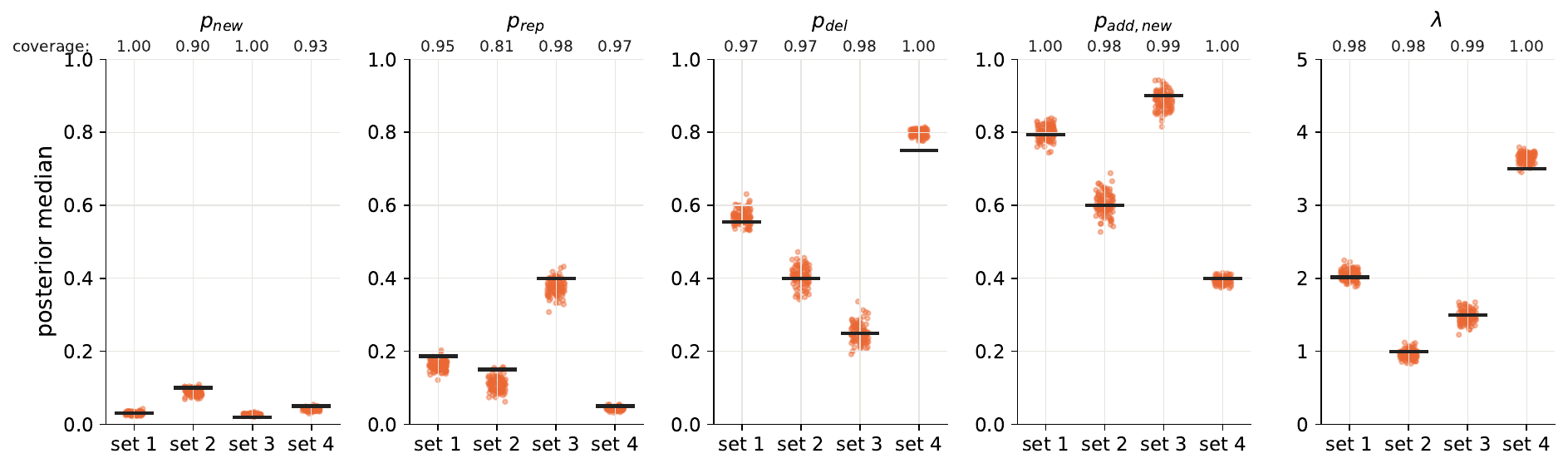}
    \caption{Recovery of known parameters from synthetic corpora. One panel per parameter and one column per parameter set (set 1 is the best fit). Black bars: true values; dots: posterior medians of $100$ corpora; numbers: coverage of the $95\%$ credible intervals.}
    \label{fig:recovery}
\end{figure*}

We generate $100$ synthetic corpora of $1317$ papers, seeded as the real one, for each of four parameter sets, $(p_{\text{new}}, p_{\text{rep}}, p_{\text{del}}, p_{\text{add,new}}, \lambda)$: the best fit (set $1$), $(0.10, 0.15, 0.40, 0.60, 1.0)$, $(0.02, 0.40, 0.25, 0.90, 1.5)$ and $(0.05, 0.05, 0.75, 0.40, 3.5)$, the last with a high turnover. To make $400$ inferences affordable, each corpus is analysed by regression-adjusted ABC~\cite{beaumont2002approximate} on a single reference table of $2\times10^5$ prior simulations. All four sets are recovered (Fig.~\ref{fig:recovery}): the $95\%$ credible intervals contain the true value in $93$--$100\%$ of the corpora in $18$ of the $20$ combinations of parameter and set. The two exceptions are $p_{\text{rep}}$ in set $2$, underestimated by about $0.03$ (coverage $81\%$), and $p_{\text{new}}$ in the same set (coverage $90\%$). In addition, ABC-SMC itself, with $400$ particles and otherwise the settings of Sec.~\ref{sec:appendix_abc}, was run on three synthetic corpora at the best fit (Table~\ref{tab:smc_spotcheck}): all $15$ credible intervals contain the true value, and the medians lie within $0.025$ of it for the probabilities and within $0.07$ for $\lambda$. With three corpora this is a consistency check rather than a coverage estimate.

\begin{table*}[tp]
\caption{ABC-SMC on three synthetic corpora generated at the best fit ($400$ particles, $20$ generations, final tolerance $\varepsilon=8.2$--$8.8$). Posterior medians with $95\%$ credible intervals; all $15$ intervals contain the true value.}
\label{tab:smc_spotcheck}
\begin{ruledtabular}
\begin{tabular}{lcccc}
 & True value & Corpus 1 & Corpus 2 & Corpus 3 \\
\colrule
$p_{\text{new}}$     & 0.031 & 0.033 [0.017, 0.049] & 0.037 [0.015, 0.055] & 0.031 [0.014, 0.050] \\
$p_{\text{rep}}$     & 0.187 & 0.202 [0.163, 0.245] & 0.171 [0.127, 0.211] & 0.198 [0.155, 0.238] \\
$p_{\text{del}}$     & 0.555 & 0.554 [0.457, 0.670] & 0.562 [0.469, 0.689] & 0.562 [0.471, 0.681] \\
$p_{\text{add,new}}$ & 0.793 & 0.772 [0.670, 0.867] & 0.797 [0.678, 0.900] & 0.817 [0.705, 0.908] \\
$\lambda$            & 2.01  & 2.08 [1.79, 2.43]    & 1.98 [1.70, 2.37]    & 1.98 [1.71, 2.32] \\
\end{tabular}
\end{ruledtabular}
\end{table*}

\section{Evidence for extending the model}
\label{sec:appendix_extensions}

In order to obtain the analytical results given in Sec. \ref{subsec:analytical}, our model is deliberately minimal: the reference paper, the protected author as well as the re-added authors are drawn uniformly at random, and the number of added authors is Poisson-distributed, following hence one of the simplest, most common probability distributions to work with.

Here we collect evidence in our data for the extensions discussed in the main text. Every statistic is computed with the same function on the data and on $50$ simulated corpora at the best fit, so that any bias of the measurement affects both equally.

\subsection{Choice of the reference paper}
\label{sec:appendix_extensions_reference}

One of the major assumptions for our model to work is the fact that every past paper is equally probable to be the reference of the one created in each time-step. This is considered like this because there is no proper way of knowing the actual reference of a modified paper, and we construct all of our summary statistics under this key premise.

\begin{figure*}[tbp]
    \centering
    \includegraphics[scale=0.8]{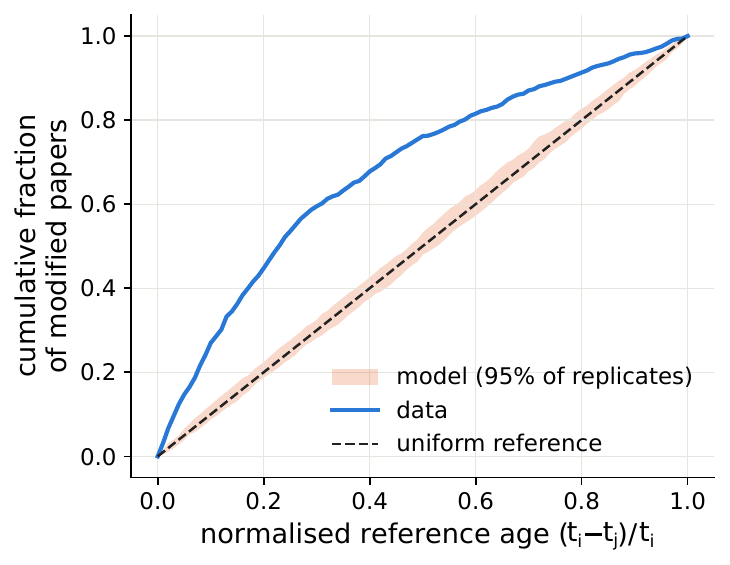}
    \caption{Cumulative distribution of the normalized age of the reference of modified papers, in the data and in simulations at the best fit ($95\%$ range). Dashed diagonal: the uniform choice of the model.}
    \label{fig:reference_lag}
\end{figure*}

To study whether there is a significant influence of publication date in the modification of a team, we decide in this subsection to consider the reference to be the earlier paper with the largest author overlap, calculated with the Jaccard index. We observe that, if references were drawn uniformly from the past, their normalized age $(t_i-t_j)/t_i$, for a paper $i$ that modifies paper $j$, would be uniformly distributed. This happens in the model, but not in the data. In fact, references in our real-case scenario are markedly recent, with median normalized age $0.23$ against $0.50$ in the model (Fig.~\ref{fig:reference_lag}). An ageing function in the choice of the reference would capture this.

\subsection{Broader distributions of added authors}
\label{sec:appendix_extensions_nb}

We decided to choose a Poisson distribution for the addition process as it would enable us to obtain analytical results for the stationary distribution of the collaboration size. However, this may fail to reflect the fact that a generated paper coming from the modification process can vastly enlarge the network---for instance, after finishing a collaboration with several people on an idea that was originally developed in some sort of conference or workshop.

For this, we replace the Poisson distribution of the number $m$ of added authors by a Poisson-gamma or negative binomial one with the same mean $\lambda$ and dispersion $r$, i.e., $P(m)=\binom{m+r-1}{m}p^{r}(1-p)^{m}$ with $p=r/(r+\lambda)$. This distribution has a variance $\lambda+\lambda^2/r$, meaning that it grows as $r$ decreases with the original Poisson distribution being the limit $r\to\infty$. We keep the rest of parameters of the model as the ones of the best fit.

\begin{figure*}[tp]
    \centering
    \includegraphics[width=\textwidth]{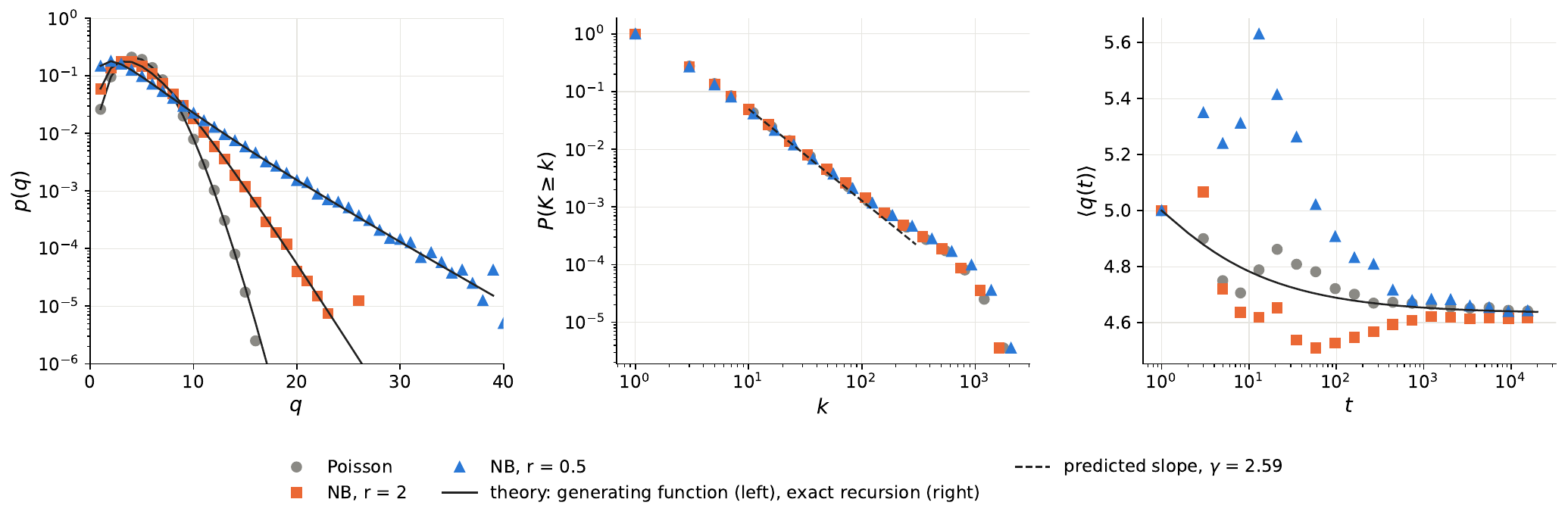}
    \caption{Poisson and negative-binomial additions with the same mean ($r=2$ and $r=0.5$), all other parameters at the best fit, grown from a single $5$-author paper to $20\,000$ papers. Left: stationary distribution of collaboration size, with the generating-function prediction (lines). Middle: complementary cumulative distribution of productivity; dashed line: slope of Eq.~\eqref{eq:gamma}. Right: average collaboration size; line: exact recursion, Eq.~\eqref{eq:exact_q}.}
    \label{fig:nb_theory}
\end{figure*}

\textit{Check of the theory.} With $r=2$ and $r=0.5$, the long-time average collaboration size and the tail of the productivity distribution coincide with those of the Poisson case, as predicted; at early times the average collaboration size fluctuates more for small $r$. In contrast, the stationary distribution of collaboration size broadens and agrees with the generating-function result of Sec.~\ref{sec:appendix_stationary_authors_papers} (Fig.~\ref{fig:nb_theory}).

\begin{figure*}[tp]
    \centering
    \includegraphics[width=\textwidth]{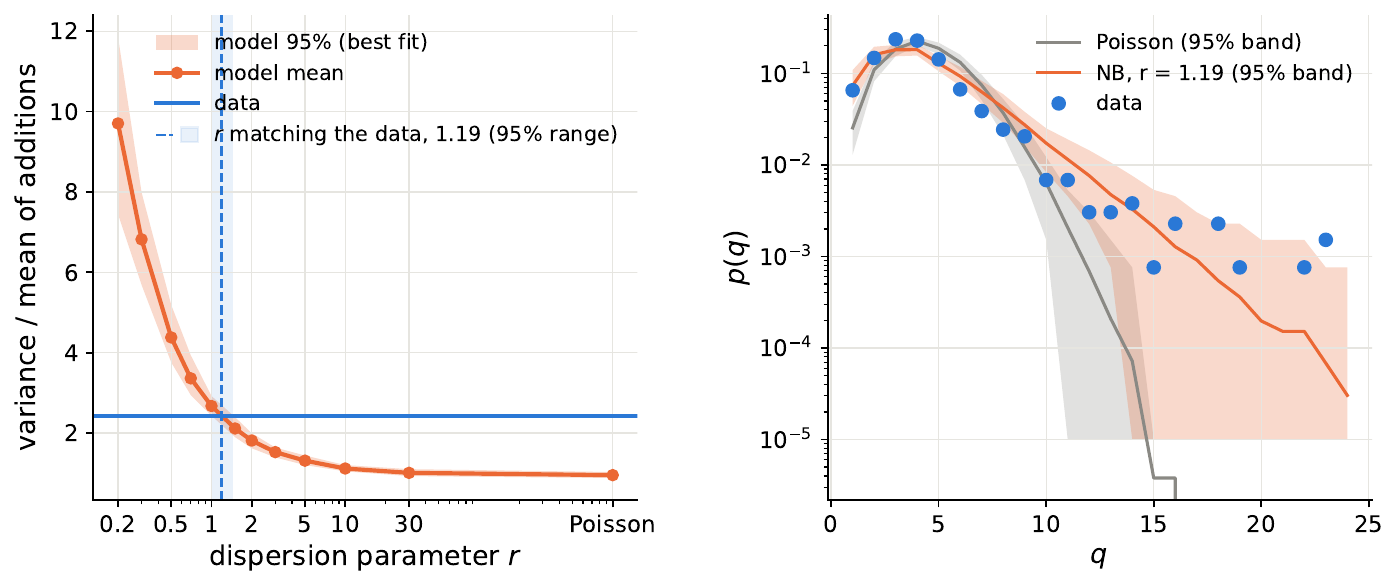}
    \caption{Left: ratio of variance to mean of the number of added authors as a function of $r$, in simulations at the best fit (mean and $95\%$ range) and in the data (horizontal line); band: values of $r$ that match the data. Right: distribution of collaboration size in the data and in simulations with Poisson additions and with $r=1.19$ ($95\%$ bands over posterior draws).}
    \label{fig:nb_data}
\end{figure*}

\textit{Comparison with the data.} In our real-case scenario, the number of authors added to a modified paper has a variance $2.4$ times its mean in the data, which is much larger than the case of having Poisson additions, where mean and variance coincide. Matching this ratio in the negative-binomial distribution gives a value for dispersion $r\approx1.2$ (Fig.~\ref{fig:nb_data}, left). With this single change of distribution, we see in Fig.~\ref{fig:nb_data}, right that the model is capable of reproducing more accurately the fraction of single-author papers, and comes much closer to that of papers with ten or more authors, which are the two main discrepancies of the distribution of collaboration size explained in the main text.

\subsection{Choice of the protected author}
\label{sec:appendix_extensions_authors}

In the model every author is treated equally, and therefore everyone has the same chance of being kept apart from the protected author when undergoing the modification process.

\begin{figure*}[tp]
    \centering
    \includegraphics[scale=0.35]{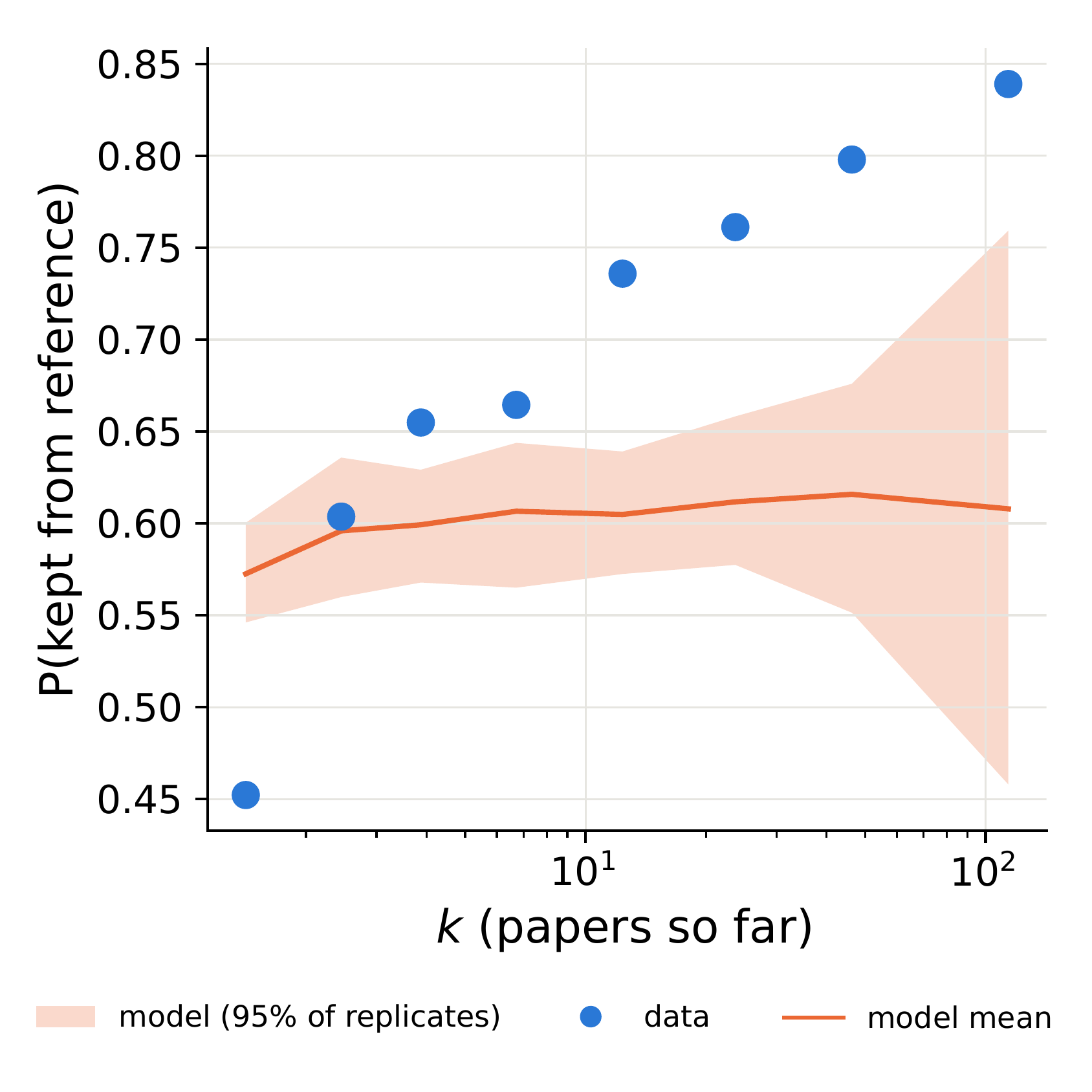}
    \caption{Probability that an author of the reference is kept in the modified paper as a function of the author's productivity, in the data and in simulations at the best fit ($95\%$ range).}
    \label{fig:retention}
\end{figure*}

Retention, however, seems to grow with the author's productivity (Fig.~\ref{fig:retention}, left) when looking at the data. This could be indicative of the idea that the most productive member of the team should be kept, as if the protected author behaved like a team leader that belongs to the institution. Therefore, a model with two populations of authors, with different probabilities of being kept and re-added, could be a the natural extension suggested by these observations.

The possibility of having each individual's role within the system would also enable more accurate predictions on members' turnover, at the expense of being more complicated to replicate among different datasets.



\label{sec:appendix_extensions_core}

\end{document}